\documentclass[fleqn,usenatbib]{mnras}

\usepackage{newtxtext,newtxmath}

\usepackage[T1]{fontenc}
\usepackage{booktabs}
\usepackage{graphicx}
\usepackage{subcaption}
\usepackage{longtable}
\usepackage{float}
\usepackage{tabularx}
\usepackage{array}
\usepackage[normalem]{ulem}

\DeclareRobustCommand{\VAN}[3]{#2}
\let\VANthebibliography\thebibliography
\def\thebibliography{\DeclareRobustCommand{\VAN}[3]{##3}\VANthebibliography}

\usepackage{graphicx}	
\usepackage{amsmath}	

\title[$T_{eff}$ and TTS in LAMOST DR10 with ML]{Effective temperatures estimation of low-mass stars and 
identification of T Tauri stars in LAMOST DR10 using machine learning}

\author[C. D. Millan-Valderrama et al.]{
C. D. Millan-Valderrama,$^{1}$ 
J. Hernandez,$^{2}$\thanks{\url{E-mail: hernandj@astro.unam.mx}} 
B. Sabogal$^{1}$, 
J. Muñoz$^{3}$ 
\\
$^{1}$, Departamento de Física, Universidad de los Andes, 
Cra 1 Nº 18A - 12, Bloque Ip, 111711, Bogotá, Colombia\\
$^{2}$Instituto de Astronom\'ia, Universidad Nacional Aut\'onoma de M\'exico, Apartado Postal 106, C. P. 22800, Ensenada, B. C., M\'exico\\
$^{3}$Facultad de Ingeniería, Universidad de los Andes, Cra. 1 Este No. 19A - 40, Bloque ML, 111711, Bogotá, Colombia
}

\date{Accepted 07/16/2026; in original form 04/14/2026}

\pubyear{\the\year{}}

\begin{document}
\label{firstpage}
\pagerange{\pageref{firstpage}--\pageref{lastpage}}
\maketitle

\begin{abstract}
We present effective temperature ($T_{eff}$) estimates of low-mass stars and T Tauri stars (TTS) candidate detections derived from automated spectroscopic measurements and low-complexity machine-learning 
models applied to the LAMOST DR10 V2 survey.
Equivalent widths 
of key diagnostic spectral features, including the atomic lines H${\alpha}$, \ion{Li}{I} $\lambda$6708{\AA} and TiO/VO molecular bands, are automatically measured for all stars within 1 kpc observed by LAMOST. Using nine spectral features as inputs, we train a Gradient Boosting Machine 
tree-based regression model, calibrated with synthetic spectra from the PHOENIX library, to predict $T_{eff}$ over the range $2,500-5,100$ K. 
We apply a logistic regression 
model to the principal components derived from the measured spectral features, enabling efficient
identification of TTS candidates. Both models exhibit strong performance in validation tests. Finally, a Monte Carlo
framework is employed to propagate input uncertainties and estimate $T_{eff}$ and its associated uncertainties for low-mass stars from 
1,733,802 spectra 
and to identify 2,534
candidate TTS from 3,121
spectra. 
\end{abstract}




\section{Introduction}

TTS represent the pre-main-sequence (PMS) phase of low-mass stars ($\lesssim 2M_{\odot}$) and are crucial for studying stellar evolution during the first tens of millions of years. Since TTS are initially surrounded by circumstellar disks of gas and dust that evolve into planetary systems \citep{ercolano2017dispersal, pietu2014faint}, they are also cornerstone objects in studies of planet formation.

TTS are generally classified into Classical TTS (CTTS) and  Weak-lined TTS (WTTS) based on the H$\alpha$ line \citep{herbig1988, white2003}. CTTS exhibit strong emission of H$\alpha$ and ultraviolet flux excesses produced by circumstellar gas accreting onto the star through magnetospheric columns that impact the stellar surface at free-fall velocities ($>$300 km/s) \citep{johns1999measuring, hartmann2016}. 
Infrared excesses, produced by the dusty disk component heated by stellar radiation, are also observed. In WTTS, accretion has ceased or has dropped below detectable levels. Consequently, WTTS exhibit H$\alpha$ emission generated by magnetic activity on the stellar chromosphere with intensities lower than those expected for CTTS of similar $T_{eff}$. Depending on their evolutionary stage, WTTS may still show infrared excesses from circumstellar disks. 
These disk-bearing WTTS may represent a transitional phase between CTTS and WTTS \citep[e.g., CWTTS;][]{briceno2019cida} or stars with very low accretion rates \citep[e.g., low-accretors;][]{Manara2013, Manara2017, thanathibodee2022}.

Another relevant spectral characteristic in TTS is the presence of the \ion{Li}{I} $\lambda$6708{\AA} absorption line (hereafter \ion{Li}{I}), which serves as a youth indicator in low-mass stars with spectral types K or M \citep{briceno2019cida,saad2024abyss, hernandez2023lamost}. Features produced by the TiO and VO molecular bands are present in this spectral type range and provide information about their $T_{eff}$ \citep{KH1995,herczeg2014optical}. The $T_{eff}$ is a fundamental stellar parameter that governs the spectral energy distribution, bolometric luminosity, and evolutionary state of a star \citep{gray2021observation}. Accurate determinations of $T_{eff}$  are essential for 
estimating stellar radii and masses, as well as for constraining atmospheric models \citep{casagrande2010absolutely, Pecaut2013}.
In addition, other emission lines associated with accretion, stellar activity, or stellar winds can appear in TTS spectra \citep[e.g., \ion{Ca}{II}, \ion{He}{I}, \ion{O}{I};][]{alcala2014,saad2024abyss}.

The current generation of large-scale, multi-wavelength surveys offers an excellent opportunity to identify TTS. 
The vast datasets generated by these surveys require efficient processing methods and tools, such as Machine Learning (ML) \citep{khalil2019big, li2025machine}. Consequently, ML applications have become increasingly prevalent in Astronomy. 

For example, in the search for PMS stars, \cite{mcbride2021untangling} developed three convolutional neural networks (CNNs) within their Sagitta framework. This approach primarily relied on photometric data from Gaia \citep{GAIADR2_2018} and the Two Micron All-Sky Survey \citep[2MASS; ][]{Cutri2003}. The first CNN generated an extinction map; the second performed a binary classification to identify PMS candidates; and the third estimated stellar ages up to $\sim$70 Myr. By cross-matching their results with LAMOST DR5 \footnote{\url{https://www.lamost.org/lmusers/}},  they measured the equivalent width (EW) of \ion{Li}{I}  for $\sim 2000$ sources with Signal-to-Noise Ratio (S/N) $> 30$ in the r-band to establish their medium-confidence threshold (probability $>85\%$). Applying their classifier to curated subsets of Gaia DR2 and EDR3 data, they identified 77,591 and 138,582 likely PMS sources (confidence $>$ 85\%), respectively. Similarly, their H$\alpha$ EW measurements showed that $84\%$ of these medium-confidence sources were consistent with being CTTS or WTTS (e.g., H$\alpha$ in emission). Finally, the authors reported that the derived ages are consistent with those of better-studied star-forming regions, while noting limitations in age determination for binary systems.

For their part, \cite{saad2024abyss} applied spectroscopic criteria to identify young stellar objects (YSO) in the Sloan Digital Sky Survey (SDSS) optical spectra (SDSS-V BOSS) with CTTS representing young stars undergoing accretion. They developed two CNNs: The first one, named LineForest, extracts 52 spectral features from a predefined list, achieving recall values of 0.96 for H$\alpha$ and 0.64 for \ion{Li}{I}. The second CNN uses 164 inputs, including these spectral features along with parameters such as $T_{eff}$, $\log g$, and photometric data from Gaia and 2MASS, to estimate YSO probabilities. Their model achieved an F1 score of 0.633 \footnote{Recall is analogous to completeness, i.e. the fraction of true objects successfully recovered by the classifier, whiles the F1-score combines completeness (recall) and classification purity (precision) into a single performance metric.}. Applying these methods they identified: $11.1 \times 10^3$ YSO candidates in SDSS-V BOSS and 6105 in LAMOST DR8, both with probabilities greater than $0.8$ of being YSO. Furthermore, analyzing the overdensity in the H$\alpha$ vs. $T_{eff}$ relation, they reported 6,444 CTTS. In particular, they observed significant \ion{Li}{I} depletion in stars of $\sim1.5-3$ Myr, followed by a plateau up to 10 Myr, after which the \ion{Li}{I} abundance continues to decline.

On the other hand, \cite{fang2025lamost} spectroscopically identified M-type YSO in LAMOST DR8. They employed two distinct approaches. In the first one, they established a threshold for the EW of \ion{Li}{I} to separate YSO from other objects, identifying 6,300 candidates. They referred to this as the ``lithium filter" method.  In the second approach, they implemented a Random Forest classifier using seven features from the so-called red band, identifying 8,750 candidates. The candidates were then categorized into four groups based on the EW of \ion{Li}{I} and H$\alpha$, obtaining 2,328 CTTS and 6,239 WTTS candidates. They reported a significant scatter in the \ion{Li}{I} measurements of YSO at different ages, which may be attributed to variability in magnetic activity.

In supervised ML methods, predictions rely on labeled samples, meaning datasets for which the properties of interest are known. Therefore, the training sample should be prepared to resemble as closely as possible the new or production sample. 
For example, all measurements used during the training and application stages of a given ML model should be derived from spectra acquired and processed in a homogeneous manner.

In the case of empirical samples, the required characteristics are more restrictive; for instance, in temperature determination, a representative sample with well-defined intervals is necessary, but not always available. 
An alternative is to use synthetic spectra as the training sample for application to empirical data, an approach that has demonstrated good performance \citep{10.1093/mnras/stac1790}. An additional advantage of this method is the ability to calibrate predicted values against a theoretical model with physical quantities directly derived from it. In this context, the PHOENIX\footnote{\url{https://phoenix.astro.physik.uni-goettingen.de/}} atmosphere code, which generates synthetic spectral libraries \citep{allard1995model, husser2013new}, is widely used for stellar parameter estimation and for the spectral analysis of stellar samples at different evolutionary stages.

The LAMOST DR10 release provides
$\sim$11.4 million low-resolution spectra, offering a valuable dataset for ML applications in determining stellar parameters and identifying TTS. In this paper, we employ EW measurements from both synthetic and empirical spectra, combined with ML techniques, to estimate $T_{eff}$ and identify TTS within 1 kpc of the Sun using LAMOST DR10. This paper is organized as follows. Section \ref{sec:Data_Featrues} describes the data samples, Section \ref{sec:MLRegClas} presents the ML methods used for regression and classification, Section \ref{sec:results} provides the results and discussion, and Section \ref{sec:summ_concl} summarizes our conclusions.

\section{DATA and FEATURES} \label{sec:Data_Featrues}

\subsection{Synthetic Spectral Dataset
} \label{sec:phoenix}

The Göttingen spectral library provides high-resolution synthetic spectra (R of 500,000 in the optical and 100,000 in the near-infrared) based on the stellar atmosphere code PHOENIX, covering the wavelength range from $0.05-5.5\: \mu$m \citep{husser2013new}. The parameter space covers $T_{eff}$ from 2300 K to 12000 K (in steps of 100 K for $T_{eff} \leq$ 7,000 K and 200 K for  $T_{eff}>$ 7000 K), surface gravity from  log($g$) 0 to 6.0 (in steps of 0.5 dex), and different metallicities ([Fe/H]) and alpha-element abundances ([$\alpha$/M]).

To characterize PMS and main-sequence low-mass stars, we selected PHOENIX spectra covering $T_{eff}$ from 2300 K to 6000 K and log($g$) from 3.0 to 5.0. Since we were 
working in the solar neighborhood, we selected spectra with [Fe/H] = 0 and [$\alpha$/M] = 0. The selected spectral library was convolved to the resolution of LAMOST (R = 1,800), covering a spectral range from 3,900 to 9,000 {\AA}. Adopting the extinction law of \citet{Cardelli1989} with $R_V$ = 3.1, we reddened the spectra using the following $A_V$ values: 0.0, 0.2, 0.5, 1.0, 1.5, 2.0, 2.5, 3.0, 4.0, and 5.0. Finally, assuming that the flux error at each wavelength can be estimated as Ferr = Flux/(S/N), we added noise to each synthetic spectrum using a Gaussian distribution for the error and the S/N values: 30, 50, 150, 200, 300, and 500. The final synthetic library includes 11,400 spectra (hereafter the adapted synthetic library, Table \ref{tab:datasets} summarizes the datasets used throughout the remainder of this work), with 38, 5, 10, and 6 different values for $T_{eff}$, log($g$), $A_V$,
and S/N, 
respectively.

\subsection{LAMOST DR10 and Gaia DR3} \label{sec:Data}

LAMOST is a multifiber spectrograph mounted on the $4$ m Schmidt telescope of the National Astronomical Observatory of China, located at the Xinglong Observing Station. This telescope includes 4000 optical fibers, 16 spectrographs, and 32 CCD cameras. Fibers are distributed over a 5{\degr} diameter field of view. In the low-resolution survey (R$\sim$1800 at 5500 {\AA}), each spectrograph contains 250 fibers with blue and red arms covering spectral ranges of 3,700–5,900 {\AA} and 5,700–9,000 {\AA}, respectively \citep{Cui2012}.

In this work, we use the database provided by LAMOST DR10, which, in addition to the spectra, includes a general catalog with basic information from the third Gaia data release (Gaia DR3), as well as S/N of the LAMOST spectra in several photometric bands (\textit{u, g, r, i, and z}), and spectral classifications from the LAMOST 1D pipeline \citep{Luo2004, Luo2012}.

For most of the spectra in LAMOST DR10, Gaia DR3 provides positions ($\alpha$ and $\delta$), parallaxes ($\pi$), proper motions ($\mu_\alpha$ and $\mu_\delta$), and photometric measurements (\textit{G}, \textit{BP}, and \textit{RP} magnitudes) with unprecedented precision and reliability \citep{creevey2023gaia}. Since the parallaxes reported by Gaia DR3 require a systematic correction, we apply the zero-point bias obtained from the ARI Gaia TAP service\footnote{\url{https://gaia.ari.uni-heidelberg.de/}}. The zero-point bias depends on the ecliptic latitude, brightness, and colors of the star \citep{Lindegren2021a}. In the following, we use the corrected parallaxes for systematics ($\pi_C$).

\subsubsection{Observed spectra datasets}
\label{sec:samples}

Initially, we required spectra with S/N $>$ 3 in the \textit{g}, \textit{r}, and \textit{i} bands, and S/N $>$ 30 in any of them. We also restricted our sample to spectra associated with Gaia DR3 sources with parallax errors better than 20\% that satisfy specific quality criteria for the following parameters: the renormalized unit weight error (RUWE), which evaluates the quality of the Gaia DR3 astrometric solution assuming a single-star model \citep{Lindegren2021b}, and the fidelity parameter, defined by \citet{Rybiski2022} using ML techniques to evaluate the quality of proper motions and parallaxes reported by Gaia DR3. The required values for these parameters were RUWE $\leq$ 1.4 or fidelity $\geq$ 0.5. 

Finally, we required sources located within 1 kpc ($\pi_C\geq$1 mas). 
This distance limit was adopted to focus on the nearby Galactic population, where low-mass PMS stars, the main targets of this work,  are better represented by spectra with adequate quality and more reliable distance estimates. Although young low-mass stars are present beyond 1 kpc, their fraction with high-quality spectra decreases with distance, making the sample increasingly affected by selection effects. In addition, this distance limit largely excludes more distant star-forming regions with potentially subsolar metallicities, avoiding the need to account for metallicity as an additional parameter in the analysis.

Of the 11,441,011 spectra in LAMOST DR10, a subset of 4,286,001 satisfied these requirements. Hereafter, we denote this subset as sample $S_{0}$. 

For the sample of known TTS, we selected stars from two star-forming regions: Taurus and the Orion Star Forming Complex (OSFC). For Taurus, we cross-correlated the $S_{0}$ sample with the catalogs of young stars reported by \citet{esplin2019survey}. For the OSFC, we cross-correlated the $S_{0}$ sample with the kinematic candidates reported by \citet{sanchez2024kinematic}.  After visual inspection to assess the presence of \ion{Li}{I} 
in absorption, H$\alpha$ in emission, and TiO or VO molecular bands in the spectra, we obtained 284 and 513 spectra in each region, respectively. 
Since one star can have more than one LAMOST DR10 spectrum, we selected one spectrum per star based on visual inspection, prioritizing the spectrum with the highest spectral quality and the clearest diagnostic features. 
This procedure avoids introducing redundant and non-independent observations that provide limited additional information on the underlying stellar properties and violate the independence assumption commonly adopted in statistical learning methods \citep{Bishop2006pattern, Goodfellow2016deep}. It also reduces the risk of information leakage during model evaluation \citep{Kapoor2023}.
Thus, the known TTS sample includes  124 stars located in Taurus and 435 stars in the OSFC.
 
The Gaia DR3 survey also provides six high-quality homogeneous samples of stars with distinct stellar and evolutionary characteristics, known as “golden samples” \citep{creevey2023gaia}. In particular, they report $\sim$3 million OBA-type stars and $\sim$3 million FGKM-type stars. We cross-correlated the $S_{0}$ sample with these two golden catalogs and randomly selected 829 spectra to construct the non-TTS sample, requiring that the spectral type reported by the LAMOST pipeline matched the expectations from the golden catalogs. 
The number of non-TTS spectra was chosen to maximize the size of the labeled sample while maintaining a reasonably balanced class distribution, with known TTS and non-TTS spectra representing approximately 40\% and 60\% of the sample, respectively. This compromise avoids introducing a severe class imbalance that could bias the learning process \citep{sun2009classification}.

\subsection{Features} \label{sec:feature}

Following the method described in \citet{Hernandez2004}, we implemented a python code \citep{hunt2019advanced} for measurements of  pseudo-EW or spectral indices obtained by measuring the decrease (in absorption) or increase (in emission) of the flux within a feature band relative to the local continuum level of the spectrum. The latter is estimated by interpolating the flux measured in two continuum bands adjacent to the feature band. These indices are largely insensitive to reddening and S/N as long as the sidebands are chosen immediately adjacent to the measured feature and the feature band is wide enough to obtain a good flux estimate \citep{Hillenbrand1995PhD,Hernandez2004}. For each feature, a judicious selection of the width and the center of each band (two sidebands and the feature band) represents a compromise between minimizing reddening effects and maximizing the S/N. To avoid large uncertainties in the measurements, we required at least values with flux greater than three times their error in each of the three bands used for estimating EW. Measurements that did not meet this requirement were reported as missing values. 
The uncertainties in the spectral indices were estimated using the MC error propagation method \citep{Anderson1976}, assuming Gaussian distributions for the flux uncertainties in the spectra and performing 300 iterations.

Based on the spectral features used by the SPTCLASS tool\footnote{\url{https://www.astrosen.unam.mx/\~hernandj/SPTclass/sptclass.html}} to classify stars across a wide range of spectral types \citep[e.g., see Table 1 in ][]{Hernandez2004}, we have measured several EW to study stars with different properties and evolutionary stages. In addition to the SPTCLASS features, we included indices optimized to detect and characterize carbon stars \citep{Jaime2026} and other indices that trace stellar activity and stellar winds. A list of spectral indices with details of the set of bands is included in the appendix \ref{apendix:bands}.

For this work, we selected features optimized to study low-mass stars (e.g., the L\_TiOb and L\_VOb bands, defined in the Table \ref{Tab:bands}). We also included the spectral indices related to the \ion{Li}{I} and the H$\alpha$ lines since they trace youth, accretion, or magnetic activity in TTS. Given the importance of these two lines in our work, we also measured EW using an automatic Python code that fits a Gaussian profile to each spectral line after normalizing the spectrum to its local continuum.

The measurement procedures were applied consistently to the adapted synthetic library and $S_{0}$ sample 
according to their specific use in the subsequent analyses. The resulting measurements constitute the feature set used throughout this work.

\begin{figure*}
    \centering
    \includegraphics[width=1.0
    \linewidth]{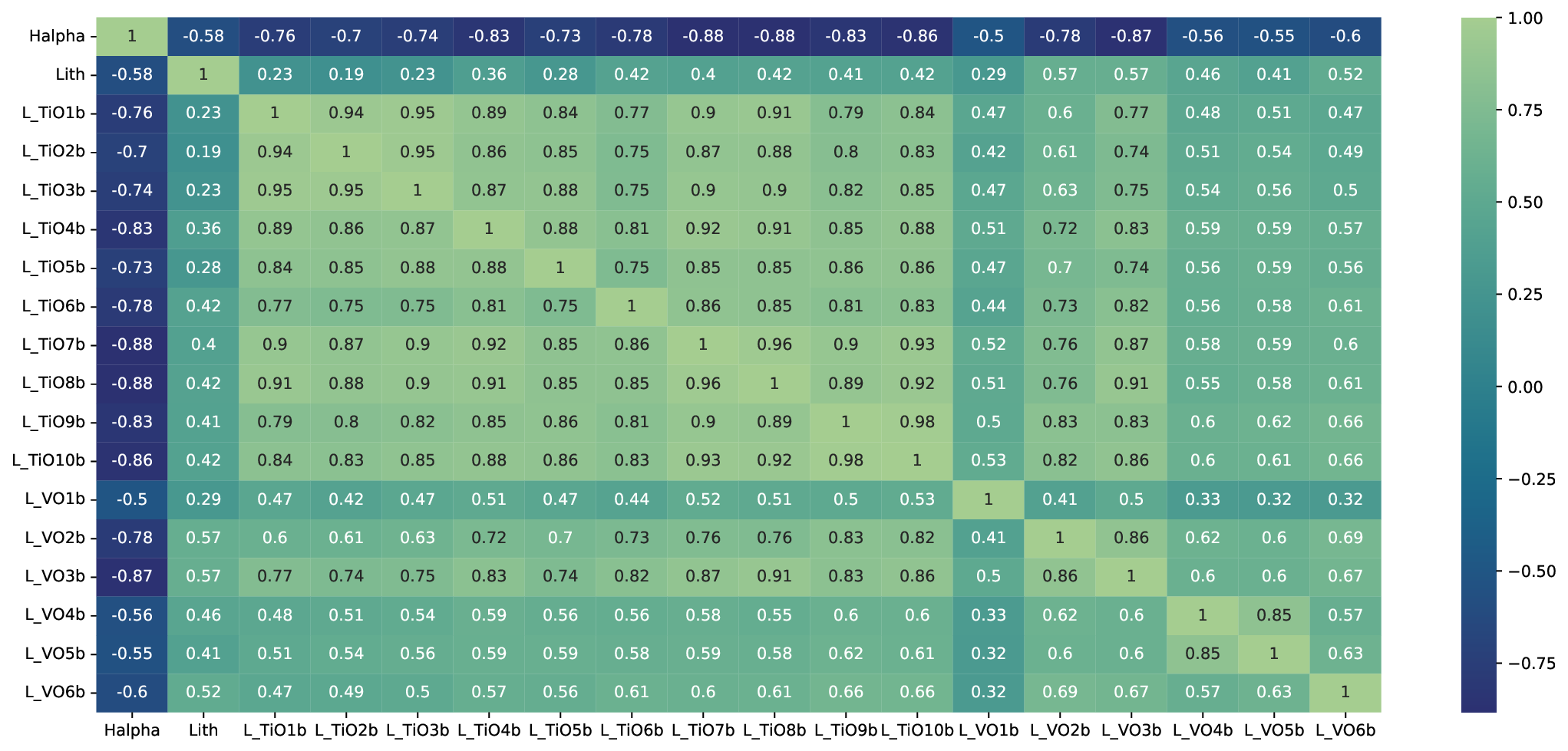}
    \caption{Correlation matrix for the 18 spectral features calculated over the 1309 sources with full measurements (485 TTS and 824 non-TTS).}
    \label{fig:corr_mat}
\end{figure*}

Figure \ref{fig:corr_mat} shows the Spearman correlation matrix \citep{Fei} for the measurements of 18 selected features in a sample of 1309 sources, corresponding to spectra with complete feature measurements, including 485 TTS and the rest classified as non-TTS. The high correlation observed among  absorption features belonging to the same molecular species is expected, as they are primarily sensitive to temperature \citep{herczeg2014optical}. A similar rationale applies to the moderately strong correlation found between the features of the two different molecular species included in this analysis.
Notably, the EW of H$\alpha$ is 
anti-correlated with the TiO and VO molecular band indices,
indicating that, in most cases, 
stronger molecular absorption 
 is associated with stronger
H$\alpha$ emission.
This suggests that stars with lower $T_{eff}$ tend to exhibit stronger H$\alpha$ emission, consistent with the scenario that low-mass stars have deeper convective zones and enhanced magnetic activity \citep{herczeg2014optical,briceno2019cida}. In addition, the largest EW of H$\alpha$ are generally observed in cooler stars owing to contrast effects, as the H$\alpha$ emission stands out against a lower photospheric continuum level \citep[e.g.,][]{Stauffer1997,Frasca2015,Manara2017}.

\section{Machine learning regression and classification models} 

\label{sec:MLRegClas}

We implemented two tools designed to automatically estimate the $T_{eff}$ of low-mass stars and to identify TTS. The first tool employs a Gradient Boosting Machine (GBM) regression algorithm, while the second uses a robust Logistic Regression (LR) classifier. A general description of each algorithm is provided below, and implementation details are presented in Section \ref{sec:results}.

\subsection{Gradient Boosting Machine}
\label{sec:GBM}

GBM is a widely used method for both regression \citep{miettinen2018protostellar}  and classification \citep{wang2024age} tasks. In regression, the goal is to predict the value of a parameter (e.g., a physical property) based on its relationship with one or more input features. In classification, the algorithm assigns instances to discrete categories based on the selected features. GBM consists of an ensemble of weak learners, typically decision trees, built sequentially. Each tree attempts to correct the errors made by the previous ones, resulting in a model with improved generalization performance.

GBM offers some advantages over other methods: It does not require feature transformation (e.g., standardization or PCA) to handle correlated inputs, it is robust to outliers, and it can handle missing data without the need for explicit imputation \citep{friedman2001greedy}. 
However, like other tree-based models, GBM has limitations when extrapolating beyond the range of the training data, and such predictions should be interpreted with caution. In this work, we implemented GBM using the H2O.ai tool \citep{h2o} that allows flexible hyperparameter tuning, efficient handling of large datasets, and integration with environments such as R, Python, and the H2O Flow Web interface.

GBM was selected because it provides an effective compromise between predictive performance, computational efficiency, and model complexity for structured tabular data. Although more advanced boosting implementations such as XGBoost are available in H2O.ai, the objective of this work was not to perform an exhaustive comparison of machine-learning algorithms, but rather to evaluate the proposed methodology using a well-established boosting approach.


\subsection{Logistic Regression}
\label{sec:LR}

LR is a classical statistical algorithm widely used for binary classification problems, and it has been broadly applied across various scientific disciplines. In LR, the probability of a given outcome (for example, whether a star is a TTS or not) is estimated by computing a weighted linear combination of the spectral features and then applying the logit function, which transforms this linear predictor into a probability bounded between 0 and 1 \citep{hosmer2013}. In astronomy, for instance, \cite{cambazard2025logistic} employed LR to enhance the detectability of exoplanets in SPHERE spectro-polarimetric images, while \cite{beitia2018use} applied it to derive membership probabilities for TTS candidates using ultraviolet and infrared photometry. Although some studies have highlighted the limitations of LR compared to more complex classifiers \citep[e.g.,][]{sharma2020classification, haghighi2022analyzing}, others have shown that LR can outperform alternative methods \citep{nusinovici2020logistic} and have emphasized the importance of proper methodology and transparent reporting when using classification models \citep{christodoulou2019systematic}.

One advantage of LR over other algorithms is its reliable performance with moderate sample sizes. However, the application of LR requires  meeting certain statistical assumptions, including  independence of observations, linearity in the logit, the absence of multicollinearity, and the absence of influential observations \citep{stoltzfus2011logistic}. When these assumptions are violated, the model may yield biased coefficients, high variance in estimators, inefficient estimates, or invalid statistical inferences, such as unreliable hypothesis tests \citep{menard2002applied}. In general, the first two assumptions are relatively easy to satisfy. Multicollinearity can be mitigated by selecting less correlated features or by applying a principal component analysis (PCA) transformation. Addressing the  influential observations may require removing (no recommended) or down-weighting these observations, or using robust estimation methods to implement a more reliable LR model \citep{maronna2019robust}.

Robust LR was adopted as a simple and computationally efficient probabilistic classifier. Following the principle of model parsimony, we selected a well-established statistical approach that reduces sensitivity to influential observations relative to the standard logistic model. Given the relatively small number of input features after dimensionality reduction, this choice allows the effectiveness of the proposed methodology to be assessed without introducing unnecessary algorithmic complexity.

\section{Results and discussions}
\label{sec:results}

Initially, we measured 16 spectral features that trace the effective temperature in low-mass stars: ten associated with TiO bands and six with VO bands (see Figure \ref{fig:corr_mat}). 
An initial assessment of missing data in the $S_{0}$ sample revealed that L\_VO6b has approximately $10\%$ missing values, while  L\_TiO4b, L\_TiO5b, L\_TiO6b and L\_VO5b, each account for approximately $1\%$ missing values. 
Among the remaining features, L\_VO4b, L\_TiO2b, and L\_TiO3b show missing values of 0.8\%, 0.4\%, and 0.4\%, respectively, and exhibit high collinearity with features that have fewer missing values.
Furthermore, the EW-$T_{eff}$ relation from the PHOENIX grid indicates that L\_VO1b is useful only for stars with $T_{eff}$$\lesssim$ 2800 K and EW $>$ 4, whereas more than 99.6\% of the $S_{0}$ sample 
shows EW values of L\_VO1b below this threshold.

In this initial evaluation of features, we explored models with a varying number of features and different levels of missing data. Although GBM can handle missing values, our exploratory results suggest that spectra with missing values in the features used by the regressor may lead to underestimated temperature predictions. This bias is particularly meaningful when the missing value corresponds to a feature of high importance.

Based on this evaluation and supported by collinearity considerations, we selected L\_TiO1b, L\_TiO7b, L\_TiO8b, L\_TiO9b, L\_TiO10b, L\_VO2b, and L\_VO3b for developing the GBM regression model (hereafter the temperature regressor; Section \ref{sec:regressor}). 
The selection of these features ensures less than ~0.4\% missing values and prioritizes the reliability and robustness of the models over the marginal contributions that might otherwise be provided.
Moreover, for the LR classifier model (hereafter the TTS classifier), we used the features L\_TiO9b, L\_TiO10b, L\_VO2b, L\_VO3b, H$\alpha$, and Lith. In this case, we required that the sample not have missing values in any of the six features. 
We also included in the LR classifier the $T_{eff}$ derived from the temperature regressor and the EW of \ion{Li}{I} derived by Gaussian fitting (gLiI), with missing values imputed to zero (see Section \ref{s:classifier}). 
In future work, we intend to investigate more general imputation techniques and alternative approaches aimed at reducing the loss of samples associated with missing feature values.

We first applied the temperature regressor to the $S_{0}$ sample, and then applied the TTS classifier to stars in the $T_{eff}$ range corresponding to spectral types K or M.

\subsection{Temperature Regressor}
\label{sec:regressor}

We used the GBM regression method to estimate the $T_{eff}$ of stars in the S$_0$ sample. To establish a theoretical calibration between the selected features and $T_{eff}$, while accounting for a range of surface gravities, noise levels, and reddening conditions, the GBM models were trained and validated using exclusively the adapted synthetic library (Section \ref{sec:phoenix}), without including the observed TTS spectra later used in the TTS classifier. The trained model was subsequently applied to the S$_0$ sample to predict $T_{eff}$. Sources with
temperature ranges for low-mass stars ($2500<T_{eff}<5100$ K) were retained for the subsequent analysis.




 
There is no general rule for partitioning data; however, the training sample should comprises between 50\% and 70\%, and the validation and testing samples should be equally divided \citep{ivezic2020statistics}. Thus, the adapted synthetic
library was split into training ($60\%$), validation ($20\%$), and testing ($20\%$) sets. The final model selection was then conducted in three main steps.

We performed a preliminary coarse exploration of GBM regression models using the seven selected features to 
assess their predictive importance and locate a promising region of the hyperparameter space for subsequent refinement.
Using k-fold cross-validation and the validation sample, and aiming to minimize model complexity, we explored a grid of models defined by the following hyperparameters: the number of trees ($ntrees =  300, 500, 700, 900$), learning rates ($learning\_rate =  0.01, 0.05, 0.1, 0.5$), and maximum tree depths ($max\_depth = 5, 10, 15 , 20$). The recommended values for k range from 5 to 10 \citep{Hastie2009}, and here we used k=5.

The 5-fold cross-validation is used to evaluate models by splitting the training sample into 5 subsets (folds). Each fold is used once as a validation set, while the remaining 4 folds are used for training; this process is repeated 5 times, and the results are averaged to estimate model performance.  

We evaluated the performance of the models using the root mean square error ($RMSE$) obtained from the 5-fold cross-validation and the validation set. The $RMSE$ value quantifies how far the model predictions deviate from the actual values, with higher values indicating larger errors. The model with the lowest $RMSE$ (hereafter the first model) has
the following hyperparameter: $ntrees = 300$,  $learning\_rate = 0.1$ and $max\_depth = 5$. 

\begin{figure}
\centering
    \includegraphics[width=1.0\linewidth]{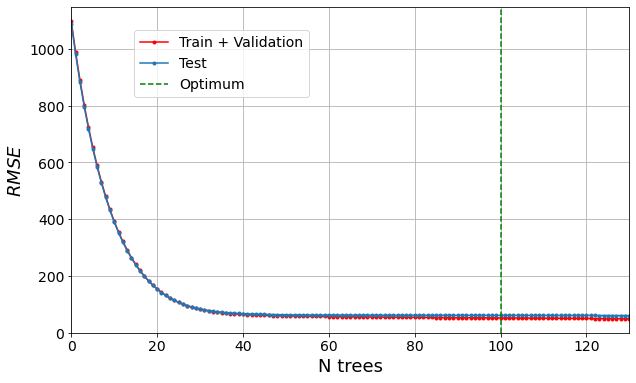}
    \label{fig:Learning_curv}

    \centering
    \includegraphics[width=1.0\linewidth]{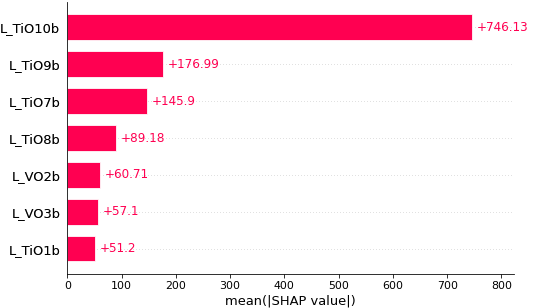}
    \caption{\textit{Top:} Learning curve for our GBM regression model to determine $T_{eff}$. \textit{Bottom:} Feature importance of our GBM regression model through of the Mean SHAP values.}
    \label{fig:Learn_curv_Impact_Shap_Val}
\end{figure}

We used the SHapley Additive exPlanations (SHAP) values to assess the feature importance \citep{lundberg2020local}. SHAP values provide a consistent measure of how much each feature contributes to the prediction by evaluating its average impact across all trees. Based on the SHAP analysis of the validation sample, we determined that all seven features have relevant importance in this first model (see the lower panel of Figure \ref{fig:Learn_curv_Impact_Shap_Val}).

Once the seven characteristics had been confirmed as relevant,
we performed a refined exploration of the hyperparameter space around the best-performing region identified in the initial search. This second stage focused on smaller tree depths and fewer trees, consistent with our objective of minimizing model complexity while maintaining predictive performance. Using all seven features, we explored
a new grid of models by combining the hyperparameters: $ntrees =  100, 200, 300$; $learning rate = 0.1, 0.5$ and $max\_depth = 3, 5 $. 
Then, based on the $RMSE$ and the coefficient of determination, $R^{2}$, obtained from the 5-folds cross-validation and validation sample, we selected the best hyperparameters.  
The coefficient $R^{2}$  measures how well a regression model explains the variance in the target variable. This is calculated by $R^2 = 1 - \frac{SS_{res}}{SS_{tot}}$, where $SS_{res} = \sum_{i=1}^{n} (y_i - \hat{y}_i)^2$ is the Residual Sum of Squares with $y_{i}$ the observed value and $\hat{y}_{i}$ is the corresponding fitted value,  and $SS_{tot} = \sum_{i=1}^{n} (y_i - \bar{y})^2$ is the Total Sum of Squares, where $\bar{y}$ is the average value \citep{montgomery2021introduction,  halswanter2022introduction}.
It ranges from 0 to 1, 
where 1 is a perfect fit, corresponding to 100\% of variance explained. 

Finally, after hyperparameter optimization, the 
model was trained using the combined training and validation subsets (80\% of the adapted synthetic sample).
This model consists of $100$ trees, a learning rate of $0.1$, and a maximum depth of $5$. The resulting model achieved $RMSE$ and $R^{2}$ values of $\sim 51$ K and $0.998$ on the combined training and validation samples,
$\sim 60$ K and  $0.997$ in 5-fold cross-validation, and $\sim 66$ K and $0.996$ in the test 
set, respectively. 
The very high $R^2$ values reflect the controlled nature of the synthetic spectra, where the relationship between the measured features and $T_{eff}$ is not affected by the additional sources of uncertainty and variability inherent to observed spectra.
The learning curves ($RMSE$ versus training iterations, i.e., as additional trees are added) for the training and test sets remain closely aligned, with no systematic divergence between training and test performance. This consistency indicates that the model generalizes well and shows no evidence of overfitting within the explored complexity range. In practice, overfitting would typically manifest as a continued decrease in training $RMSE$ while test $RMSE$ stalls or increases as model complexity grows. The upper panel of Figure \ref{fig:Learn_curv_Impact_Shap_Val} illustrates the learning curve for 130 trees, suggesting that approximately $100$ trees represent  
the optimal number.

To estimate uncertainties in $T_{eff}$ from the flux errors of the spectra, we applied a Monte Carlo (MC)-based error propagation approach \citep{Anderson1976}, assuming that the uncertainties in the TiO and VO bands, as estimated in Section \ref{sec:feature}, follow Gaussian distributions. For each spectrum, we generated 500 $T_{eff}$ predictions by varying the EW of the TiO and VO bands within their respective error bars. The mean and standard deviation of these 500 predictions were adopted as the estimated $T_{eff}$ and its associated uncertainty ($\mathrm{e}T_{eff}$), respectively.

Following the spectral type – effective temperature conversion table of \citet{Pecaut2013}, we selected as K- and M-type stars those whose spectra led to estimated $T_{eff}$ values between 2,500 and 5,100 K. The total number of spectra within this temperature range is 1,733,852. Hereafter, we refer to them as the Low-Mass Star sample (LMS sample). Table \ref{Tab:LMS_sample} lists the $T_{eff}$ values and other relevant parameters for the LMS sample. 

\begin{table}
\centering
\caption{Basic source information, spectral feature measurements and estimated quantities of the LMS sample.}
\label{Tab:LMS_sample}
\begin{tabular}{@{}lcl@{}}
\toprule
Column            & Unit              & Description                                                   \\ \midrule
obsid             &                   & LAMOST Unique spec. ID for each observation \\
gaia\_source\_id  &                   & The ``source\_id" field of Gaia DR3 catalog                    \\
RA                & deg               & Right Ascension in J2000                                      \\
DEC               & deg               & Declination in J2000                                          \\
Subclass          &                   & LAMOST subclass                                               \\
Sample           &                   & prod: production; labeled: effective labeled \\
L\_TiO1b          & \AA               & EW of TiO1b                      \\
eL\_TiO1b         & \AA               & EW uncertainty of TiO1b           \\
L\_TiO7b          & \AA               & EW of TiO7b                      \\
eL\_TiO7b         & \AA               & EW uncertainty of TiO7b           \\
L\_TiO8b          & \AA               & EW of TiO8b                        \\
eL\_TiO8b         & \AA               & EW uncertainty of TiO8b           \\
L\_TiO9b          & \AA               & EW of TiO9b                      \\
eL\_TiO9b         & \AA               & EW uncertainty of TiO9b           \\
L\_TiO10b          & \AA               & EW of TiO10b                        \\
eL\_TiO10b         & \AA               & EW uncertainty of TiO10b           \\
L\_VO2b          & \AA               & EW of VO2b                      \\
eL\_VO2b         & \AA               & EW uncertainty of VO2b           \\
L\_VO3b          & \AA               & EW of VO3b                      \\
eL\_VO3b         & \AA               & EW uncertainty of VO3b           \\
Lith              & \AA               & \ion{Li}{I} EW from the bands method      \\
e\_Lith           & \AA               & \ion{Li}{I} EW uncertainty from the bands method                     \\
Halpha            & \AA               & H$\alpha$ EW from the bands method                            \\
e\_Halpha         & \AA               & H$\alpha$ EW uncertainty from the bands method                \\
gLiI               & \AA               & \ion{Li}{I} EW  from the Gaussian fit                                \\
e\_gLiI            & \AA               & \ion{Li}{I} EW uncertainty from the Gaussian fit                     \\
n\_gLiI            &                   & Iterations used for the gLiI measurement            \\
gHa               & \AA               & H$\alpha$ EW  from the Gaussian fit                                \\
e\_gHa            & \AA               & H$\alpha$ uncertainty from the Gaussian fit                     \\
n\_gHa            &                   & Iterations used for the gHa measurement            \\
$T_{eff}$         &  K & Effective temperature                             \\
e\_$T_{eff}$&  K& Uncertainty Effective temperature                             \\ 
lo\_p             &                   & Lower CI limit of the TTS probability                         \\
p                 &                   & TTS probability                                               \\
up\_p             &                   & Upper CI limit of the TTS probability                         \\
p$_{thr}$             &                   &   specific probability threshold                      \\
flag\_TTS         &                   & Flag for TTS candidate selection                  \\
flag\_CTTS        &                   & Flag for CTTS classification based on Halpha                  \\
flag\_gCTTS       &                   & Flag for CTTS classification based on gHa                    \\
flag\_GaiaCMD        &                   & Flag for stars younger than 50 Myr in the CMD      \\

\hline
\multicolumn{3}{l}{(This table is available in its entirety in machine-readable form.)}\\
\end{tabular}
\vspace{1pc}
\end{table}

\subsubsection{Comparison with large-scale stellar parameter catalogues}
\label{s:Teff_largescale}

In the LMS sample, we compare the results derived from our GBM regressor with those reported by LAMOST DR10, Gaia DR3, and the $T_{eff}$ estimates from APOGEE-Net \citep{olney2020apogee,sprague2022} and BOSS-Net \citep{Sizemore2024,saad2024abyss}, which apply 
CNN methods in near-infrared and optical spectra, respectively.

\begin{figure*}
    \centering
    \includegraphics[width=0.9\linewidth]{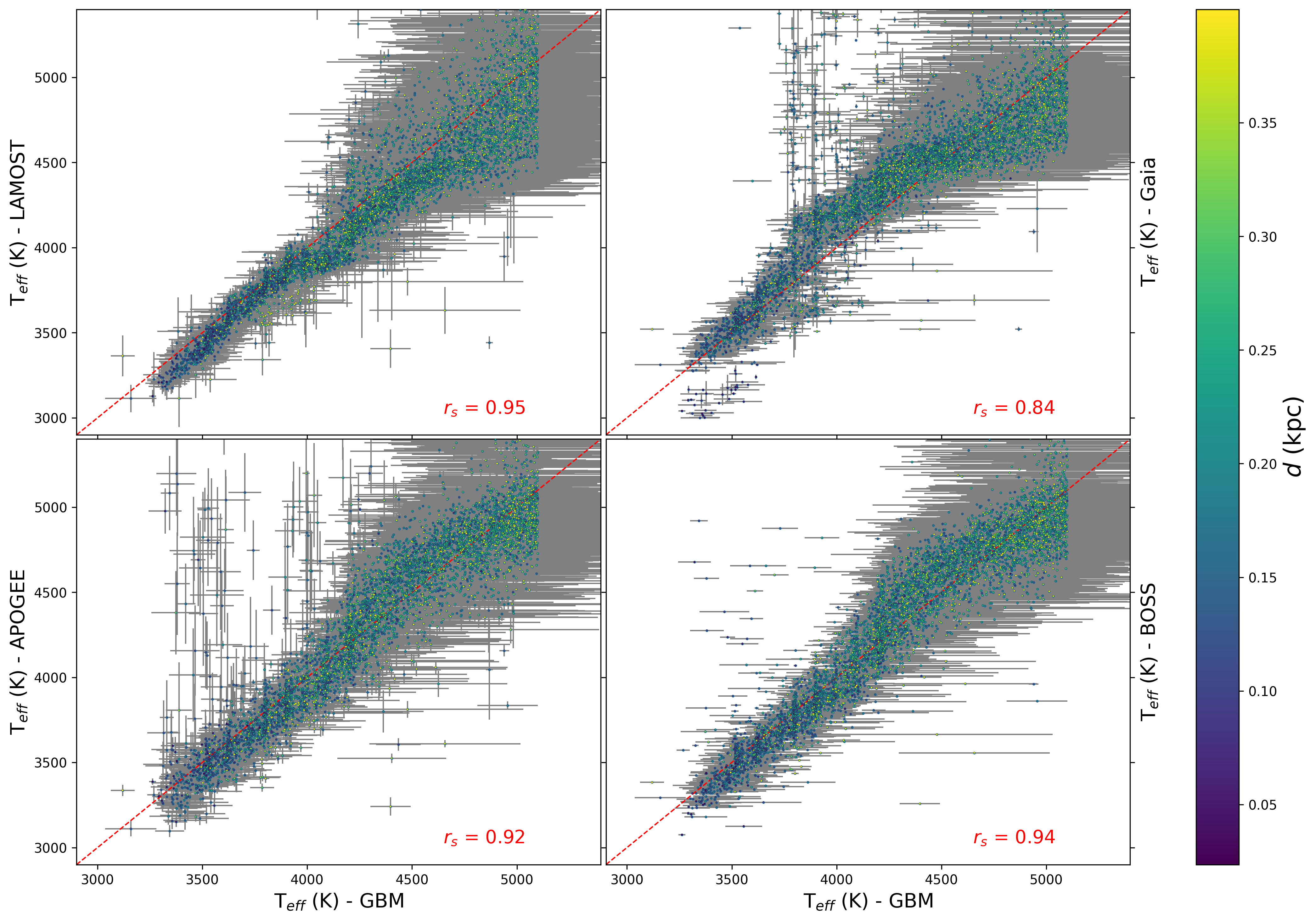}
    \caption{ Comparison of the $T_{eff}$ within 0.4 kpc for around 6,300 common sources of the catalogs LAMOST (top left panel), Gaia (top right panel), APOGEE (bottom left panel) \citep{olney2020apogee}, and BOSS (bottom right panel) \citep{Sizemore2024} with these obtained by our GBM regression model. The Spearman  correlation coefficient is indicated in each inset.}
    \label{fig:Scatter_Teffs}
\end{figure*}

The LAMOST Stellar Parameter pipeline \citep[LASP;][]{Wu_Du_Luo_Zhao_Yuan_2014} derives stellar atmospheric parameters from the flux-calibrated one-dimensional spectra produced by the LAMOST 1D pipeline. LASP operates in two stages. First, it performs a coarse search over a grid of synthetic and empirical templates to obtain initial estimates of $T_{eff}$, surface gravity, metallicity, and radial velocity. In the second stage, these parameters are refined through a $\chi^2$-minimization procedure, using interpolated spectra to achieve an optimal match with the observations. Starting from the sixth LAMOST data release, \citet{Du2021LASPM} presented an optimized version of LASP, specifically designed to improve stellar parameter estimates for M-type stars. In a small number of cases, the reported LASP $T_{eff}$ and their associated uncertainties are set to negative values, which may indicate problems with the estimates. Specifically, 2,351 spectra in the LMS sample have negative values in LASP $T_{eff}$. In addition, 390,914 spectra ($\sim$ 33\%) of the LMS sample lack an estimated LASP $T_{eff}$. 

Using GSP-Phot \citep{andrae2023gaia}, Gaia assigns $T_{eff}$ by fitting the low-resolution \textit{BP}/\textit{RP} spectra and Gaia broad-band photometry (\textit{G}, \textit{BP}, \textit{RP}) with synthetic spectral libraries covering a wide range of stellar parameters. With a Bayesian inference framework, GSP-Phot derives $T_{eff}$, surface gravity, metallicity, and extinction simultaneously, selecting the solution that best reproduces the observed spectral energy distribution while accounting for interstellar reddening. There is a temperature–extinction degeneracy \citep{andrae2023gaia}, which can be particularly problematic in star-forming regions \citep{hernandez2023lamost}.
Gaia DR3 report $T_{eff}$ for 1,462,999 ($\sim$84\%) spectra in the LMS sample.

The APOGEE-Net tool \citep{olney2020apogee, sprague2022} estimates $T_{eff}$ using a CNN trained on APOGEE near-infrared spectroscopic data. The network was designed to model a wide range of stellar types, including low-mass MS stars, red giants, PMS stars, and OB-type stars \citep{sprague2022}. During training, their labeled samples were drawn from previous APOGEE results and other catalogs with known stellar parameters, allowing the model to learn spectral feature patterns across a broad parameter space. Generally, CNN estimations and uncertainties depend on the quality and representativeness of the labels; they can fail when applied to data outside the training domain \citep{shen2025engression}.
From the LMS sample, 235,999 spectra have a $T_{eff}$ estimated from APOGEE-Net.

\citet{Sizemore2024} developed a complementary CNN model, BOSS-Net, for optical BOSS (Baryon Oscillation Spectroscopic Survey) spectra, aiming to replicate the performance of APOGEE‑Net on these optical data through label transfer.
This CNN was trained using theoretical stellar models and stars with labeled optical spectra from BOSS and LAMOST. From the LMS sample, 1,522,541 spectra have a BOSS-net $T_{eff}$.
 
Figure \ref{fig:Scatter_Teffs} shows a comparison between our GBM-estimated values and those from the catalogs mentioned above, for a subsample of 6,300 stars in common across those catalogs. The Spearman correlation coefficients indicate generally good agreement across the four comparisons. The best agreement is obtained between GBM and the LASP method, which is expected, as it was principally designed for LAMOST data.

Although we find relatively good agreement between GBM and GSP-Phot, some discrepant objects remain, which could reflect the temperature–extinction degeneracy inherent to the GSP-Phot method. The larger error bars and increased dispersion in the GBM estimates at higher temperatures are naturally produced by the smaller EW values expected for the TiO and VO bands, resulting in larger relative errors in the measurements of the features used by the GBM regressor.
In addition, the $T_{eff}$ estimates derived by the regressor do not account for stellar multiplicity, non-photospheric contributions (e.g. veiling or infrared excesses), or thermal surface inhomogeneities (e.g. stellar spots), and are less reliable in the $\sim4500 - 5100$ K range (see Section \ref{s:Teff_caveats}).
In future versions of the temperature regressor, we will incorporate spectral features sensitive to solar-type stars (e.g., $T_{eff} > 5200$ K), which should improve the precision, thereby reducing the error bars, while further improving the overall accuracy 
at the hot end of Figure \ref{fig:Scatter_Teffs}.

\subsubsection{Comparison with PMS-focused spectroscopic temperature estimates}
\label{s:Teff_PMS}

\begin{figure}
    \centering
    \includegraphics[width=\linewidth]{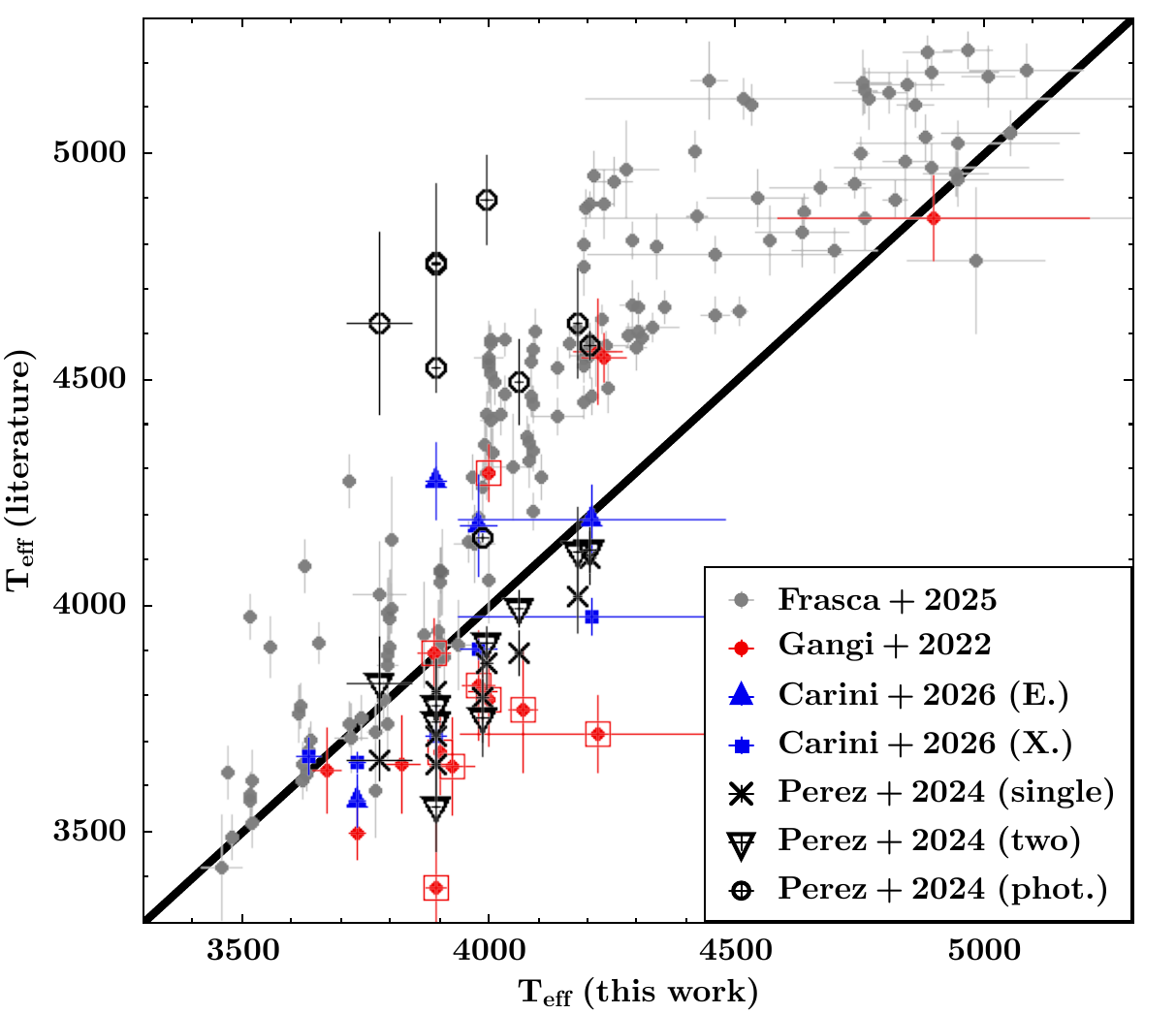}
    \caption{
    Comparison between $T_{eff}$ derived with our GBM regressor and values reported in dedicated studies of PMS stars. Grey circles correspond to \citet{Frasca2025}, red diamonds to \citet{Gangi2022}, and blue symbols to \citet{Carini2026}, using ESPRESSO (triangles) and X-Shooter (squares) spectra. Red open squares indicate $T_{eff}$ reported by \citet{Gangi2022} using a two-temperature approach, corresponding to the photospheric and spot temperature components. Black symbols indicate temperatures from \citet{Perez2024}, including single-temperature estimates (crosses) and results from a two-temperature approach, where inverted triangles correspond to the derived $T_{eff}$ values and circles represent the photospheric temperature component. The solid line indicates the one-to-one relation.}
    \label{fig:Teff_PMS}
\end{figure}

In the previous section, we compared the $T_{eff}$ derived from our GBM regressor with estimates from large-scale stellar parameter catalogues spanning different evolutionary stages. Here, we compare our results with studies of PMS stars based on high-resolution spectroscopy, broader wavelength coverage, or explicit treatments of spotted photospheres and accretion-related veiling in CTTS. Figure \ref{fig:Teff_PMS} compares our $T_{eff}$ estimates with those reported in selected PMS-focused studies. For stars with more than one GBM-derived temperature, we used the mean value after propagating the individual uncertainties through an MC error-propagation method \citep{Anderson1976}.

\citet{Carini2026} analyzed spectroscopic data obtained as part of the PENELLOPE program \citep{Manara2021} using the ESPRESSO, UVES, and X-Shooter instruments, covering optical to near-infrared wavelengths and spanning spectral resolutions from (R$\sim$17,500) to (R$\sim$140,000). They estimated $T_{eff}$, log(g), projected rotational velocity, and veiling ($r_\lambda$) using the ROTFIT tool \citep{Frasca2015}, which derives stellar parameters by minimizing $\chi^2$ between observed spectra and template libraries over selected spectral regions. We have four and five sources in common with their ESPRESSO and X-Shooter samples, respectively. Although the overlap is small, we tentatively note that differences between $T_{eff}$ estimates derived from different instruments in \citet{Carini2026} are sometimes comparable to, or larger than, the differences between their X-Shooter temperatures and our GBM-derived values. This behaviour may reflect the greater sensitivity of bluer spectral regions to stellar spots and accretion veiling \citep[e.g.][]{Manara2021, Flores2022, Gangi2022, Frasca2025}.

\citet{Frasca2025} derived stellar parameters from LAMOST mid-resolution spectra of stars in the Pleiades using ROTFIT to simultaneously estimate $T_{eff}$, log(g), projected rotational velocities, and radial velocity. Their analysis revealed discrepancies between temperatures estimated from the blue and red spectral arms, which they interpreted as a possible consequence of cool stellar spots. We have 162 stars in common with their sample. Figure \ref{fig:Teff_PMS} shows systematic differences in the $\sim4000-4500$ K interval, a regime where previous studies have reported near-infrared temperature diagnostics systematically lower (by up to $\sim$500 K) than optical estimates in spotted PMS stars \citep[e.g.][]{Flores2022,Gangi2022, Perez2024}. The comparison with temperature estimates explicitly accounting for stellar spots  \citep[e.g.][]{Gangi2022, Perez2024} suggests that part of the discrepancy observed by \citet{Frasca2025} may be related to spot-induced thermal inhomogeneities.

\citet{Gangi2022} derived $T_{eff}$, log(g), radial velocity, rotational projected velocities, and $r_\lambda$ from HARPS-N spectra of stars in Taurus–Auriga using ROTFIT. For heavily spotted stars, they adopted a two-component temperature approach in which the photospheric temperature ($T_{\rm phot}$) was fixed to the one-temperature ROTFIT estimate, while the spot temperature ($T_{\rm spot}$) was varied to reproduce the observed spectra through a composite photosphere+spot model. Similarly, \citet{Perez2024} derived $T_{eff}$ for T Tauri stars in Taurus–Auriga from SpeX–IRTF spectra ($\sim0.70-2.55$ $\mu{\rm m}$) using two-component spectral models that additionally included a blackbody contribution from the inner disk to reproduce the near-infrared continuum. Unlike \citet{Gangi2022}, \citet{Perez2024} treated both $T_{\rm phot}$ and $T_{\rm spot}$ as free parameters within an MCMC framework and also reported single-temperature estimates for comparison. In these two-component approaches, the effective temperature is estimated according to: $T_{eff}^{4}=T_{\rm phot}^{4}(1-f_{\rm spot})+T_{\rm spot}^{4}f_{\rm spot}$, where $f_{\rm spot}$ is the spot-filling factor.

Figure \ref{fig:Teff_PMS} shows that our $T_{eff}$ estimates are generally consistent with the spectroscopic determinations reported by \citet{Gangi2022} and \citet{Perez2024}, although systematic differences are present depending on the adopted methodology. For the full \citet{Gangi2022} sample, we find a weighted mean offset of $72 \pm 28$ K and a weighted $RMS$ scatter of $\sim241$ K. Restricting the comparison to the heavily spotted subsample analysed with a two-temperature approach increases the weighted mean offset to $\sim180$ K, whereas the non-spotted subsample shows a substantially smaller offset of $\sim32$ K. By contrast, the comparison with \citet{Perez2024} yields weighted $RMS$ scatters below $70$ K and weighted mean offsets of $\sim97$ K and $\sim150$ K for their two-temperature and one-temperature approaches, respectively.

The comparatively improved agreement with \citet{Perez2024}, particularly for their two-temperature approach, may reflect methodological differences relative to \citet{Gangi2022}. In the latter study, the spotted-star analysis adopts a photospheric temperature fixed to the one-temperature ROTFIT estimate, while only the spot temperature is varied. In contrast, \citet{Perez2024} allow both $T_{\rm phot}$ and $T_{\rm spot}$ to vary and include an inner-disk contribution when modelling the near-infrared continuum, resulting in a more flexible description of spotted TTS spectra. The use of near-infrared spectra, where molecular absorption bands are more prominent, may also contribute to the comparatively better agreement with our estimates. Although the infrared excess associated with the inner disk mainly affects wavelengths longer than $\sim1\:\mu{\rm m}$, beyond the dominant spectral range used by our regressor, the comparatively smaller offsets and scatters obtained with the \citet{Perez2024} two-temperature model are consistent with this approach providing more robust estimates of $T_{eff}$. Nevertheless, we note that the comparisons with \citet{Gangi2022} and \citet{Perez2024} are largely restricted to stars with $T_{eff}\lesssim4300$ K and relatively small samples. Therefore, a broader comparison spanning a wider temperature range and larger PMS samples will be required to establish a more comprehensive assessment of the agreement between different methodologies.

Within the temperature range and PMS samples explored here, the agreement observed in Figure \ref{fig:Teff_PMS} is encouraging. A possible explanation is that the most influential spectral features of our GBM regressor (see lower panel of Figure \ref{fig:Learn_curv_Impact_Shap_Val}), particularly the TiO-band indices around $\sim7000 - 7200$ \AA, lie in a spectral region where the contrast between star-spots and the stellar photosphere is smaller than at bluer wavelengths and where the excess veiling continuum associated with accretion and inner-disk emission is reduced \citep{Fischer2011,Perez2024}. In addition, our methodology relies on EW 
measurements defined relative to a local continuum, which may reduce sensitivity to broad continuum distortions introduced by veiling, infrared excesses, or thermal surface inhomogeneities, compared with approaches based on global spectral fitting. However, this interpretation remains tentative and should be tested through dedicated simulations and multiwavelength analyses of spotted and accreting PMS stars.

\subsubsection{Caveats and limitations of the temperature regressor}
\label{s:Teff_caveats}

The GBM-based temperature estimator adopted in this work presents several advantages. First, the predicted $T_{eff}$ values are derived using only seven spectral features that are largely insensitive to reddening and to broad continuum variations introduced by instrumental calibration effects. Second, the regressor is anchored to the theoretical PHOENIX spectral library, allowing the model to be applied to spectroscopic surveys with similar spectral coverage and resolution. Finally, the methodology is relatively straightforward to interpret, as it establishes a correspondence between a limited set of physically motivated spectral indices and the inferred stellar temperature.

Despite these advantages, the temperature regressor is subject to the following caveats and limitations, which should be considered when interpreting the derived $T_{eff}$ values:

\begin{itemize}
\item \textit{Model dependence}. The regressor is calibrated using the PHOENIX synthetic spectral library and therefore inherits the underlying assumptions, physical prescriptions, and approximations adopted in these models. Systematic differences arising from the use of alternative stellar atmosphere libraries are not explored in this work and remain beyond its scope.

\item \textit{Applicability range}. The temperature estimates become less reliable toward the hot end of the adopted calibration range ($\sim4500 - 5100$ K). In this regime, the TiO- and VO-band features used by the regressor weaken and become less sensitive to temperature changes, increasing the uncertainties in the inferred $T_{eff}$ values. For this reason, temperatures derived in this interval should be interpreted with caution.

\item \textit{Single-star assumption and unresolved multiplicity}. The regressor assumes a single stellar photosphere and does not explicitly account for unresolved binaries or multiple systems. Composite spectra arising from unresolved companions may bias the measured spectral indices and therefore the inferred temperature. A proper treatment of stellar multiplicity would require additional information on the stellar components and orbital configuration (e.g., orbital phase at the time of observation), which is beyond the scope of this work.

\item \textit{Non-photospheric flux contributions}. In young stars, particularly CTTS, the observed spectra may include additional continuum contributions associated with accretion processes (e.g. veiling) or irradiated dust in the inner disk \citep[e.g., infrared excesses;][]{Manara2021, Gangi2022, Perez2024}. These effects may alter the apparent strength of the spectral features used by the regressor. Although the most influential spectral indices of our model (Figure \ref{fig:Learn_curv_Impact_Shap_Val}) lie near to $\sim7000 - 7200$ \AA, a spectral region where veiling and disk-related continuum excesses may be reduced relative to bluer or near-infrared wavelengths, such contributions should still be regarded as a source of systematic uncertainty in the inferred $T_{eff}$.

\item \textit{Thermal surface inhomogeneities}. The regressor assumes a single-temperature stellar photosphere and therefore does not explicitly account for cool spots or other surface thermal inhomogeneities \citep[e.g.,][]{Flores2022, Gangi2022, Frasca2025,Perez2024}. Within the limited temperature range explored in the comparisons of Section \ref{s:Teff_PMS}, approaches that explicitly account for stellar spots generally yield systematic differences of order $\lesssim$150 K relative to our estimates.
Such approaches typically require additional free parameters (e.g. spot temperature and filling factor) and broad wavelength coverage extending into the near-infrared. Moreover, because spot visibility is modulated by stellar rotation, the observed spectrum and inferred $T_{eff}$ may vary with rotational phase \citep{Perez2024}. Therefore, stellar spots should be regarded as a potential source of systematic uncertainty in the temperatures derived by our regressor.

\item \textit{Dependence on data quality and feature measurements}. The reliability of the inferred $T_{eff}$ values depends on the quality of the measured spectral features. Low S/N, imperfect continuum placement, missing features, bad pixels, cosmic rays, or instrumental artifacts may affect the EW 
measurements used as model inputs, increasing the uncertainty or bias in the derived temperatures.

\end{itemize}


\subsection{TTS classifier}
\label{s:classifier}

The labeled sample initially contains 559 TTS and 829 non-TTS sources. However, in practice, the effective sample size available for training a ML model depends on the fraction of missing feature values in the production sample. 

Our final model uses the following features: L\_TiO9b, L\_TiO10b, L\_VO2b, L\_VO3b, H$\alpha$, Lith, gLiI, and $T_{eff}$, which together exhibit less than 1\% missing values in the LMS sample. We discarded from the training sample all spectra with missing values in any of these seven features.
We also included the EW of \ion{Li}{I} measured by Gaussian fitting (gLiI) as an additional feature. Since the Gaussian fit cannot be reliably performed when the \ion{Li}{I} line is too weak and comparable to the noise, the missing gLiI values were reasonably imputed as zero. Thus, the effective labeled sample consists of 552 TTS and 822 non-TTS sources. 

For our LR model, we split the labeled sample into training (80\%) and testing (20\%) sets.
To mitigate the effects of strong correlations among input features (Figure \ref{fig:corr_mat}), the eight features were first standardized using robust scaling; then, we applied cosine kernel PCA to the scaled features, as implemented in Scikit-learn \citep{scikit-learn} which resulted in dimensionality reduction. The choice of the kernel and the number of principal components were treated as hyperparameters and optimized during model training.  The model was trained using two principal components, which together explained 92.3\% of the variance. The influential observations in the LR model was addressed using  the robust estimation method of \cite{bianco1996robust}, as implemented in the Robustbase package in R \citep{croux2003implementing}.

In this configuration, the model demonstrates excellent class separation, consistently assigning near-zero probabilities to instances of the non-TTS class. Based on this behavior, a probability threshold of 0.1 was selected as a lower decision boundary. This value lies well outside of the non-TTS class distribution, ensuring robust classification and minimizing potential contamination from non-TTS instances. The probabilistic accuracy of the model was further assessed using the Brier score \citep{rufibach2010use} in the test set, which yielded a value of 0.015. The Brier score measures the mean squared difference between predicted probabilities and actual binary outcomes, serving as a key indicator of calibration and predictive precision. Lower scores reflect better-calibrated models, with a score of 0 indicating perfect accuracy.
The overall model performance was evaluated using the selected probability threshold, confirming the reliability and discriminative ability of the classifier under the specified configuration.

\begin{figure*}
    \centering

    \begin{subfigure}[b]{0.63\textwidth}
        \centering
        \includegraphics[width=\textwidth]{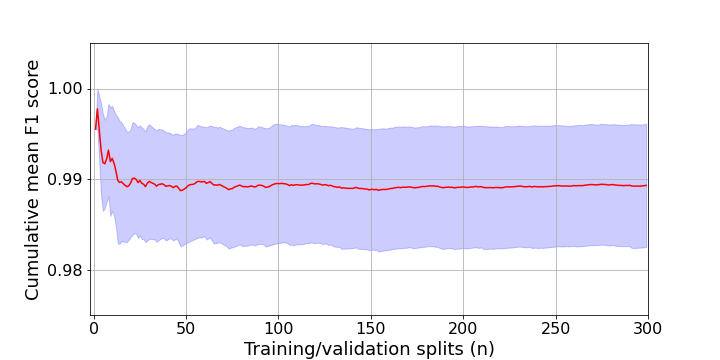}
    \end{subfigure}
    \hfill
    \begin{subfigure}[b]{0.33\textwidth}
        \centering
        \includegraphics[width=\textwidth]{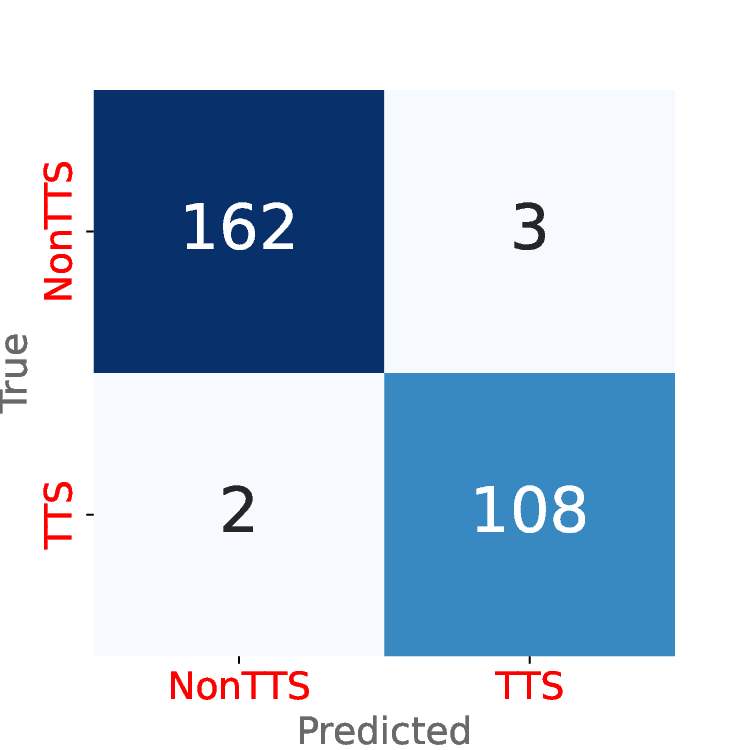}
    \end{subfigure}
    
    \vspace{0.5cm} 
    
    \begin{subfigure}[b]{0.4\textwidth}
        \centering        
        \includegraphics[width=0.6\textwidth]{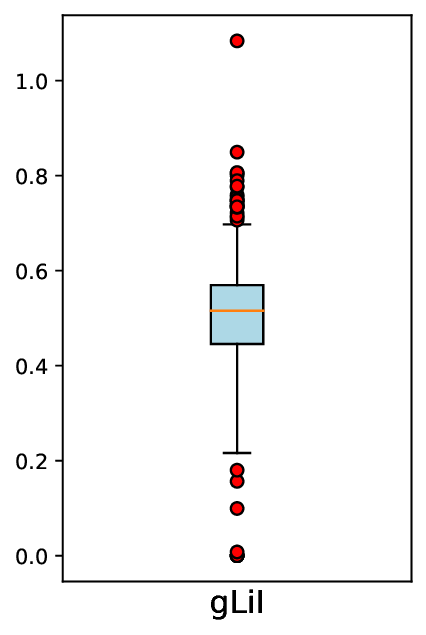}
    \end{subfigure}
    \hfill
    \begin{subfigure}[b]{0.5\textwidth}
        \centering
        \includegraphics[width=1.0\textwidth]{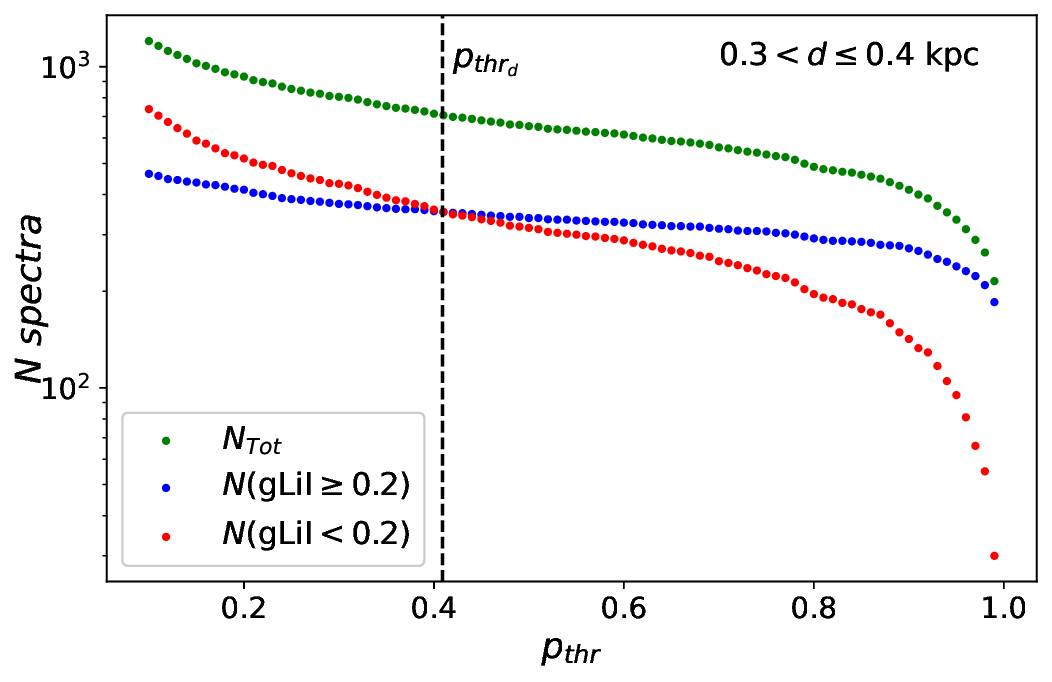}
    \end{subfigure}
    
    \caption{\textit{Top left:} Model stability assessment: cumulative mean F1-score with $\pm$ 1 standard deviation band over 300 random training/validation splits. \textit{Top right:} Confusion matrix for test sample for our LR classification model. With only 3 false positives and 2 false negatives among the 275 test samples, the model demonstrates a balanced and robust classification performance for both TTS and non-TTS objects.  \textit{Bottom left:} Adjusted boxplot for the gLiI distribution for TTS in training sample.     \textit{Bottom right:} Example for selection of $p_{thr_d}$ in samples binned of 0.1 kpc.}
    \label{fig:ModStab_ConfMat_F1max}
\end{figure*}

To evaluate the robustness and stability of the model, the training sample was repeatedly and randomly partitioned to create validation subsamples of the same size as the testing sample (i.e., 20\% of the effective labeled sample). As for the test sample, we generated 1,000 MC realizations for each validation configuration by perturbing the feature values within their measurement uncertainties, assuming a Gaussian error distribution. For each split, the validation sets were robustly scaled and kernel PCA-transformed using the parameters derived from the corresponding training subsample. A robust LR model was fitted to each training subset. The probability outputs for each spectrum were defined as the 10th percentile of the probability distribution obtained from the 1,000 MC realizations, and the model was evaluated on the corresponding validation subset using the F1 score for the class TTS. The cumulative mean and standard deviation of the F1-score were then computed over 300 iterations to quantify the stability of the model performance under repeated resampling. The results are shown in the top-left panel of Figure \ref{fig:ModStab_ConfMat_F1max}, where the standard deviation of the cumulative mean F1 score was approximately $6.80 \times 10^{-3}$ after 300 iterations.


The final model achieved in the test set a recall of $0.98$ and an F1 score of $0.98$, as indicated by the confusion matrix in the top-right panel of Figure \ref{fig:ModStab_ConfMat_F1max}. From the confusion matrix, we observe only 3 false positives and 2 false negatives out of 275 test samples, confirming that the model maintains a balanced and robust performance across both classes (TTS and non-TTS). Other performance metrics (e.g., precision, specificity, and accuracy) can also be derived from the confusion matrix, although the metrics reported here are sufficient to characterize the model performance. In addition, we assessed the goodness of fit using the Hosmer--Lemeshow test \citep{Hosmer1980}, obtaining a test statistic of 5.04 with a $p$-value of 0.75. This test evaluates whether the probabilities predicted by the LR model, obtained after applying the inverse logit (sigmoid) transformation to the linear predictor, are consistent with the observed binary outcomes. A high $p$-value indicates no evidence against the null hypothesis of adequate fit, meaning that the predicted probabilities align well with the observed data \citep{hosmer2013}.
As an additional check, we evaluated calibration with the Ebrahim-Farrington goodness of fit test \citep{farrington1996assessing} implemented in the R package \citep{ebrahim.gof2025}. The test yielded  a z‑statistic of -0.36 with $p$-value of 0.64, providing no evidence against the null hypothesis of adequate fit. This result is consistent with the Hosmer-Lemeshow test and supports that the predicted probabilities by our LR model are broadly compatible with the observed binary outcomes across the range of fitted values.

In the production stage, the trained classifier was applied to the production sample (1,732,478 spectra), derived from the LMS sample by removing the effective labeled sample.

Similarly, the predicted probability for each new spectrum in the production sample was defined as the 10th percentile of the 1,000 probability realizations. In addition, a 95\% confidence interval (CI) was computed using a bootstrap method \citep{murphy2022probabilistic}, based on the 2.5\% and 97.5\% percentiles. 
The  bootstrap procedure was implemented by resampling with replacement 500 times from the 1,000 probability estimates for each source.

The variability of new data within the production sample generally requires selecting a specific probability threshold ($p_{thr}$) that may differ from the one used for the test sample. 
Determining this threshold can be challenging because it is both model-dependent and sensitive to the properties of the input data, and it can be estimated using different approaches. 
For example, \cite{he2024identification} adopted a fixed $p_{thr} = 0.99$, whereas \citet{mcbride2021untangling} argued that the optimal $p_{thr}$ should be chosen based on the distribution and characteristics of the production sample. 
Here, we adopted both approaches.

We computed a new distance-dependent threshold by estimating a $p_{thr_d}$ in 0.1 kpc bins. Using the minimum whisker of the adjusted boxplot \citep{hubert2008adjusted} for the distribution of the gLiI feature for TTS in the training sample, $\sim$0.2 {\AA} corresponds to the lowest value still considered a non-outlier (see the bottom-left panel of Figure \ref{fig:ModStab_ConfMat_F1max}). Thus, the total number of classified spectra can be expressed as $N_{Tot} = N(\text{gLiI}\geq 0.2)+ N(\text{gLiI}<0.2)$. When the two terms in this expression are comparable, that is, when the contribution of  $N(\text{gLiI}<0.2)$ to $N_{Tot}$ becomes subdominant, the corresponding value of $p_{thr}$ can be considered the optimal threshold for that distance bin (see the bottom-right panel of Figure \ref{fig:ModStab_ConfMat_F1max}). If this condition is not satisfied, the lower decision boundary threshold of the model ($p_{thr}=0.1$) is adopted instead.

In the production process, we applied the same requirements to the production sample as to the training sample. Specifically, we required complete feature values for all seven selected features \citep{Hastie2009}, and the missing gLiI values were imputed as zero.  Spectra with incomplete feature values were excluded during the classification stage, because both the kernel PCA transformation and the subsequent robust LR classification require complete input features.
Thus, TTS classification probabilities ($p$) were computed for 1,719,380 
spectra of the production sample, hereafter referred to as the classified sample. The resulting probabilities were then added to Table \ref{Tab:LMS_sample} with the specific probability threshold ($p_{thr}$).  Spectra with probabilities  $ p \geq p_{thr}$ were classified as TTS spectra. In addition, as a measurement-quality filter, we required the gLiI EW to be recovered in at least 90\% of the MC iterations.  Using this criterion, we identified 3,121 
TTS candidate spectra, corresponding to 2,534 
unique sources, which define the TTS candidate spectra sample.

Finally, we applied the criterion of \citet{saad2024abyss} to separate CTTS and WTTS, based on the distribution of H$\alpha$ as a function of $T_{eff}$. This criterion relies on the study of \citet{white2003}. Specifically, for the $T_{eff}$ values included in the LMS sample, the limit between CTTS and WTTS is described by a second-order polynomial: 
\begin{equation*}
EW_{H\alpha} = -436.95 \times \log(T_{eff})^2+3227.7 \times \log(T_{eff})-5963.2.    
\end{equation*}

\begin{figure}
    \centering
    \includegraphics[width=\linewidth]{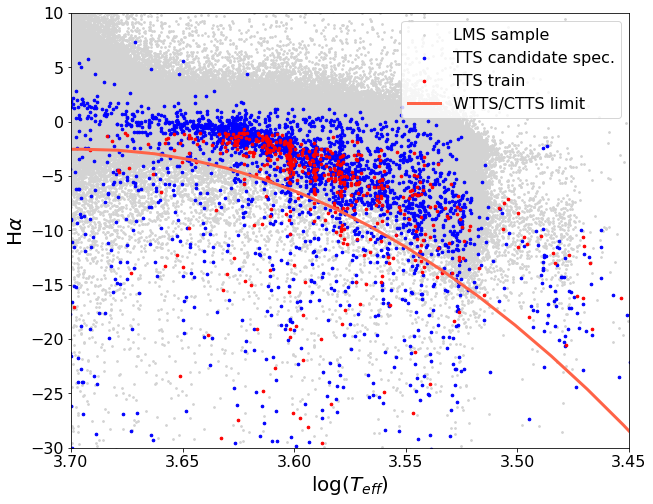}
    \caption{Separation boundary between CTTS and WTTS. This figure is similar to the \protect\cite{saad2024abyss} that stablishes a limit through H$\alpha$ as a function of log($T_{eff}$) for separate the star populations CTTS and WTTS.}
    \label{fig:TTS_criterionSaad}
\end{figure}

Figure \ref{fig:TTS_criterionSaad} shows the relation between $T_{eff}$ and the EW of H$\alpha$ for the LMS sample, indicating the CTTS/WTTS limit. In Table \ref{Tab:LMS_sample}, we flagged 809 
spectra in the TTS candidate spectra sample  
as CTTS. Additionally, 218
spectra associated with the TTS sources in the effective labeled sample 
meet the criterion for CTTS. 
Spectra in both samples 
that do not 
meet this criterion were flagged as WTTS. 
Figure \ref{fig:TTS_criterionSaad} also shows a subtle vertical pattern, mainly between log($T_{eff}$) = 3.60 and 3.55. This pattern was previously noted and is primarily associated with the discrete $T_{eff}$ values present in the training sample used for the GBM regression model.

\subsubsection{Validation of classified TTS}

In this section, we compare our results for the TTS candidate spectra sample 
with those reported in other studies. In general, the performance of the classifier depends not only on the training process but also  on the variability and quality of the production sample. Because young stars such as TTS exhibit significant observational and spectral variability, some degree of misclassifications is  unavoidable. In addition to the different distances of the stars in the sample, which affect the S/N of the spectra, some instrumental effects may also be present. LAMOST applies a sky-subtraction correction using an oversampled 2D supersky constructed from the sky fibers in each field of 20 square degrees \citep{Zhu2012}. A background gradient caused by moonlight scattering on bright nights could affect the accuracy of the sky-subtraction during the LAMOST reduction process \citep[e.g.,][]{Bai2017}. This issue can also occur in star-forming regions, where variable background emission from molecular clouds may introduce additional uncertainties. Moreover, since LAMOST fibers have an angular diameter of 3.3\arcsec, signals from more than one star can fall within the fiber apertures in crowded regions, producing blended spectra \citep[e.g.,][]{Fu2022}.
A flux discontinuity may also appear at the junction between the blue and red arms, caused by the separate optical paths and response curves of each arm \citep{Du2016}. Imperfect fiber centering, variable seeing, or calibration errors can introduce these artificial flux jumps; however, 
their impact can be mitigated by 
using 
EW measurements instead of 
the full spectral continuum.

Table \ref{Tab:Comparison} shows the number of TTS reported in previous works and the corresponding number of spectra and associated sources in LAMOST DR10. We also include 
the number of spectra and associated sources that meet our quality criteria (see Section \ref{sec:samples}) and the number of spectra and associated sources in the LMS sample (see Section \ref{sec:regressor}). From these latter numbers, we list the spectra and associated sources in the classified and TTS candidate spectra sample. 

Based on their photometric analysis and CNNs classification, \citet{mcbride2021untangling} reported 138,582 Gaia DR3 sources with a probability (p\_PMS) greater than 0.85 of being PMS stars. Of these sources, 2,317
are included in the classified sample. 
About 42.8\% (992
spectra) were recovered by our TTS classifier, a percentage that increases to 
50.6\% 
when we consider stars with p\_PMS $\geq$ 0.90. Since the method of \citet{mcbride2021untangling} relied primarily on photometric rather than spectroscopic data, we expect a higher contamination rate in their TTS classification.

\citet{saad2024abyss} reported 5,114 spectra in LAMOST DR8 with p\_PMS $>$ 0.5 and $T_{eff}$ within the LMS sample range ($T_{eff} \lesssim$  5100 K). About 2,724
spectra are in our classified sample. 
From this sample, we recovered 59.7\% (1,625
spectra), a percentage that increases to 74.7\% when the probability threshold is set to p\_PMS = 0.90. \citet{saad2024abyss}  used 164 input features, including Gaia and 2MASS photometric data. Although our classifier is based on relatively few spectroscopic features, making it computationally efficient and more interpretable, it achieves a high recovery rate of TTS compared to the more complex model of \citet{saad2024abyss}.

Using a Random Forest classifier on LAMOST DR8, \citet{fang2025lamost} reported 8,567 TTS (2,328 CTTS and 6,239 WTTS). Of the 5,857 
spectra in our classified sample, 
we recovered a 27.6\% (1,619 
spectra). We note that their training sample presents a strong induced class imbalance: only about 8\% (881 spectra) correspond to M-type YSO, whereas the remaining spectra consist of 5,048 M dwarfs and 4,868 M giants. This imbalance is not explicitly discussed in their paper, and the resulting model may therefore be more sensitive to the majority classes \citep{sun2009classification}. Additionally, they used overall accuracy as the primary evaluation metric, which is not optimal in such an imbalanced scenario; a more suitable choice would be the macro-averaged F1 score. Thus, the reported TTS sample may include a non-negligible level of contamination.

Table \ref{Tab:Comparison} also includes comparisons with TTS samples identified using alternative approaches. For example, \citet{olivares2023cosmic} applied a Bayesian methodology to identify 1,052 kinematic members in the Perseus star-forming region. Of the 226 spectra from their dataset that are included in our classified sample, 
we recovered 66.8\% (151 spectra).

In addition, \citet{zhang2024young} reported 657 YSO by analyzing Zwicky Transient Facility (ZTF) light curves, Gaia, WISE, and 2MASS photometry, and LAMOST databases. Of the 169
spectra in our classified sample,
we recovered 72\% (123 
spectra).

We selected 1,029 
spectra (corresponding to 909 sources) from the TTS candidate spectra sample that were not included as YSO in the references listed in Table \ref{Tab:Comparison}. 
About 645 spectra are not associated with a source listed in the SIMBAD database \citep{SIMBAD} and may represent previously unreported TTS candidates. After visual inspection of this selection 
we have confirmed 588 spectra (57.1\%) as TTS. 
Most of the remaining candidates correspond to spectra affected by strong [\ion{S}{II}] ($\lambda \lambda$6716, 6731\AA) contamination near the \ion{Li}{I} line, background-subtraction residuals, low S/R, or instrumental artifacts such as flux discontinuities, bad pixels, and cosmic rays. These effects may either hinder the visual confirmation of genuine TTS or lead to false-positive classifications; therefore, the rejected candidates should not be interpreted simply as poor-quality spectra.

\begin{table*} 
\centering
\caption{The values in each column indicate the numbers (N.) of spectra and sources, respectively. 
N. by auth.: these reported by the authors; 
N. in DR10: these found in LAMOST DR10; 
N. in $S_{0}$: these contained in the $S_{0}$ sample; 
N. in LMS: these in the LMS sample, the number of our labeled TTS is shown in parentheses; 
N. in class. s.: these in the classified sample; 
N. mtd.: these that matched to the TTS candidate spectra sample;
and \% mtd.: percentage of matched objects 
relative to the classified sample.}
\label{Tab:Comparison}
\begin{tabular}{@{}lccccccccc@{}}
\toprule
Author                       & N. by auth. & N. in DR10     & N. in $S_{0}$  & N. in LMS   			& N. in class. s.   & N. mtd.     	& $\%$ mtd.   \\ \midrule 
\cite{mcbride2021untangling} & 138,582     & 4,856 - 3,318	& 3,096 - 2,307  & 2,745(340) - 2,096 	& 2,317 - 1,760	& 992 - 776     & 42.8 - 44.1   \\
\cite{saad2024abyss}         & 9,375       & 9,147 - 6,323	& 3,857 - 3,152	 & 3,348(525) - 2,705 	& 2,724 - 2,213 & 1,625 - 1,326 & 59.7 - 59.9   \\
\cite{fang2025lamost}        & 8,567       & 11,197 - 8,046	& 6,486 - 5,211  & 6,466(440) - 5,203 	& 5,857 - 4,756	& 1,619 - 1,246 & 27.6 - 26.2   \\
\cite{olivares2023cosmic}    & 1,052       & 581 - 281 		& 317 - 184 	 & 231(0) - 142 	  	& 226 - 141		& 151 - 102     & 66.8 - 72.3   \\
\cite{zhang2024young}        & 342         & 473 - 330 		& 210 - 165		 & 196(22) - 155	  	& 169 - 135   	& 123 - 102     & 72.8 - 75.6   \\

\bottomrule
\end{tabular}%
\vspace{1pc}
\end{table*}

Finally, we combined Gaia DR3 parallaxes and photometry to estimate the contamination level $C$ in our
TTS candidate spectra sample. Following \citet{sanchez2024kinematic}, we fitted a linear approximation to the low-mass range ($B_P - R_P > 1.0$) of the 50-Myr PARSEC isochrone \citep{Marigo2017}. Sources below this approximation in the Gaia color-magnitude (CMD) diagram were considered potential contaminants according to:
\begin{equation*}
R_{P} - 5 \log(1000/\pi_C) + 5 < 2.1 (B_{P} - R_{P}) + 3.0.
\end{equation*}

A total of 601 
spectra were considered potential old contaminants, implying a contamination level of  $C=19.3\%$. 
For comparison, \citet{saad2024abyss} and \citet{mcbride2021untangling} report lower contamination levels of
$C=1.0\%$ and 
$C=3.1\%$, respectively. 
These lower values are expected, as their selection methods rely on features similar to those used in our contamination estimate.
In contrast, the TTS sample reported by \citet{fang2025lamost}, 
which is primarily based on spectroscopic features, shows a higher contamination level of 
$C=31.5\%$.

Furthermore, following an approach similar to that of \citet{fang2025lamost} to assess potential contamination, we compared 
the TTS candidate spectra sample with 741 members of the relatively old clusters Pleiades, M34, Praesepe, and Hyades compiled by \citet{fang2018stellar}, with ages of 125, 220, 650, and 605 Myr, respectively. A total of 2,129 spectra in our production sample are associated with stars in these catalogs, and 98.5\% of them do not meet our criteria for classification as TTS spectra. We further compared our sample with that of \citet{meingast2021extended}, who reported 7,925 sources belonging to 10 open clusters (ages $\sim30-300$ Myr) within  $ d< 500$ pc. Of the 2,012 spectra in our production sample associated with sources in this catalog, 97.3\% were not classified as TTS by our classifier. The very low fraction of TTS classifications among members of these older open clusters indicates a very low contamination rate ($<3\%$), further supporting the reliability and discriminative power of our classification method.

\subsubsection{Caveats and limitations of the TTS classifier}

Although the LR classifier presented in this work provides an efficient
framework for identifying TTS candidates from a limited set of spectroscopic features, several caveats and limitations should be considered when interpreting the derived classifications.

\begin{itemize}

\item \textit{Dependence on the training sample.} As in any supervised machine-learning approach, the classifier performance depends on the representativeness of the labelled sample used for training. The present model was trained using bona fide TTS and non-TTS samples with observational properties characteristic of nearby star-forming regions. Consequently, sources with substantially different properties from those represented in the training sample may be less efficiently recovered. In particular, distant TTS observed with lower S/N, or spectra affected by stronger observational uncertainties, may show weaker or noisier spectral signatures, potentially reducing the reliability of the classification.

\item \textit{CTTS and WTTS assignment.} The distinction between CTTS and WTTS is not derived from a dedicated machine-learning model, but from the application of an empirical H$\alpha$ EW 
criterion as a function of $T_{eff}$. Therefore, uncertainties in the measured H$\alpha$ EW
and inferred $T_{eff}$ values may propagate into the CTTS/WTTS assignment.

\item \textit{Contamination from active stars and observational effects.} Chromospherically active binaries (e.g. BY Dra and RS CVn systems) may exhibit enhanced H$\alpha$ emission and rotationally broadened spectral features that partially mimic accretion signatures commonly associated with TTS. In addition, imperfect sky subtraction, variable nebular background emission in star-forming regions, blended spectra within the $3.3^{\prime\prime}$ LAMOST fiber aperture, and instrumental artefacts (e.g. flux discontinuities, bad pixels, or cosmic rays) may affect the EW 
measurements of H$\alpha$, \ion{Li}{I}, and molecular bands used by the classifier. We note that 
of the flagged  as CTTS spectra, only 
20 are classified by LASP as 
cataclysmic variable (CV) or double star (DoubleStar), suggesting that contamination from such sources is limited, although it cannot be completely excluded.
\end{itemize}

\section{Summary and Conclusions}
\label{sec:summ_concl}

Based on 4,286,001 LAMOST spectra matched to Gaia DR3 sources within 1 kpc and satisfying the quality criteria described in Section \ref{sec:samples}, we established a spectroscopy-driven methodology that combines automated measurements with low-complexity ML 
models to estimate $T_{eff}$ through GBM regression and to identify TTS candidates through robust LR classification.

The GBM regression model was trained using the PHOENIX theoretical grid and seven spectral indices from TiO and VO molecular bands. The model achieved high predictive accuracy in $T_{eff}$, with an $RMSE$ of $\sim 66$ K  and $R^{2} = 0.996$ on the test sample. 

An MC approach was adopted to estimate $T_{eff}$ and its associated uncertainty. Assuming Gaussian uncertainties in the spectral indices, we generated 500 realizations and computed $T_{eff}$ as the mean of the resulting predictions, with the uncertainty given by their standard deviation.

Applying this model to the entire dataset yielded 1,733,852 spectra related to sources with $2,500 \leq T_{eff} \leq 5,100$ K, defining our LMS sample. The predicted $T_{eff}$ values show a strong Spearman correlation with those reported in LAMOST DR10 ($r_{S}=0.95$), demonstrating that the model effectively maps synthetic spectral indices to their empirical counterparts.

The LR classifier was trained on 552 confirmed TTS in the Taurus and Orion star-forming regions and 822 field stars selected from the Gaia DR3 golden sample. The model incorporates four molecular-band indices, the GBM-derived $T_{eff}$, and measurements of the \ion{Li}{I} and H$\alpha$ lines. 
Dimensionality reduction through kernel PCA was applied, and the final classification relies on two principal components. The model achieved an F1 score of $0.98$ and a Brier score of $0.015$.
Probabilities were defined as the 10th percentile of 1,000 MC realizations, with confidence intervals derived from bootstrap resampling. Applying the model to the LMS sample, we identified 3,121 TTS candidate 
spectra, corresponding to 2,534 
unique sources, 
Among the subset of newly identified candidates subjected to visual inspection, 588 spectra (57.1\%) were confirmed as TTS, providing an additional empirical validation of the classification methodology.

Our contamination analysis indicates a rate of approximately 20\% based on the Gaia color–magnitude diagram. However, comparisons with samples of stars in older open clusters (ages $>$ 30 Myr) reduce the estimated contamination to $\sim3 \%$, suggesting that photometry-based estimates may overestimate the contamination level. 
We also find a generally good recovery rate compared to TTS identified by more complex models that use both spectroscopic and photometric features.
In particular, we recovered $\sim 60\%$ of the TTS reported by \citet{saad2024abyss}.

Overall, our methodology provides a scalable and interpretable framework for stellar parameter estimation and PMS classification in large spectroscopic surveys. The identification of additional diagnostic features in other spectral ranges would allow the extension of the models to a broader $T_{eff}$ interval, while the incorporation of complementary photometric information represents a promising direction for further improving classification performance.


\section*{Acknowledgements}

The authors would like to thank the anonymous referee for valuable comments and suggestions that improved our earlier version of this manuscript. C D Millan-Valderrama acknowledges financial support from Fondo Doctoral para Pasantías de la Facultad de Ciencias de la Universidad de los Andes, Colombia, through the Programa de Investigación código INV-2023-178-2996. J. Hernández acknowledges financial support from the projects UNAM-DGAPA-PAPIIT IG101723 and IN110126. We thank Prof. Alejandro García Varela for valuable discussions and comments that contributed to the early development of this work. We thank the Activity and Rotation of Young Stellar Objects (ARYSO) collaboration for their valuable discussions and contributions. Guoshoujing Telescope (the Large Sky Area Multi-Object Fiber Spectroscopic Telescope LAMOST) is a National Major Scientific Project built by the Chinese Academy of Sciences. Funding for the project has been provided by the National Development and Reform Commission. LAMOST is operated and managed by the National Astronomical Observatories Chinese Academy of Sciences. This work presents results from the European Space Agency (ESA) space mission Gaia. Gaia data are being processed by the Gaia Data Processing and Analysis Consortium (DPAC). Funding for the DPAC is provided by national institutions, in particular the institutions participating in the Gaia MultiLateral Agreement (MLA).

\section*{Data Availability}

The LAMOST DR10 V2 spectra used in this work are publicly available from the LAMOST survey archive. Gaia DR3 data were accessed via the ESA Gaia Archive. The PHOENIX synthetic spectra used for model calibration are publicly available. The dataset generated in this work is available on Zenodo under persistent DOI: \href{https://zenodo.org/records/19502707?preview=1&token=eyJhbGciOiJIUzUxMiJ9.eyJpZCI6ImI1YWFkZThmLTM4ZGUtNDFhYy04Y2VmLTc3YmNhMGI2ODIzYyIsImRhdGEiOnt9LCJyYW5kb20iOiJlMzVmYTc5ZTZiOGU0MWZkN2Q2YjU2Nzc1ZmY2NjQ3MSJ9.EeqjX41oC4asMP6Vp7bK0bYIThk8toouzlezyj7Ifi4sAqgBVdCowDuM4NR0pxM2P-9ZpB-ievT--4ZTra9ykQ}{10.5281/zenodo.19502707}.
 



\bibliographystyle{mnras}
\bibliography{Bibli} 




\clearpage
\appendix
\section{Spectroscopic Indices} \label{apendix:bands}

Band configuration for the mesured spectral features are given in Table \ref{Tab:bands}.

\begin{table}
\centering
\caption{Band configuration for each spectral feature. We include the center ($\lambda$) and the width ($\Delta$) for the Central Feature Band (FB), the adjacent Blue Continuum Band (BCB) and Red Continuum Band (RCB).}
\label{Tab:bands}
\footnotesize
\setlength{\tabcolsep}{4pt}
\renewcommand{\arraystretch}{1.1}
\begin{tabular}{@{}lcccccc@{}}
\hline
Name & $\lambda_{FB}$ & $\Delta\lambda_{FB}$ & $\lambda_{BCB}$ & $\Delta\lambda_{BCB}$ & $\lambda_{RCB}$ & $\Delta\lambda_{RCB}$ \\ \hline
\hline
CaIIK     & 3933         & 20                 & 3915          & 15                      & 3949              & 12                  \\
HeI       & 4026         & 6                  & 4014          & 12                      & 4032              & 6                   \\
MnI       & 4032         & 6                  & 4026          & 6                       & 4038              & 6                   \\
FeI       & 4047         & 6                  & 4039          & 6                       & 4054              & 8                   \\
Hdel      & 4102         & 40                 & 4053          & 8                       & 4159              & 8                   \\
HeI2      & 4144         & 10                 & 4136          & 6                       & 4159              & 15                  \\
CN        & 4175         & 25                 & 4152          & 15                      & 4197              & 10                  \\
TiII      & 4176         & 6                  & 4156          & 20                      & 4196              & 6                   \\
TiII2     & 4203         & 6                  & 4183          & 10                      & 4209              & 6                   \\
CaI       & 4226         & 10                 & 4210          & 8                       & 4235              & 8                   \\
FeI2      & 4271         & 10                 & 4263          & 6                       & 4281              & 6                   \\
Gband     & 4300         & 20                 & 4285          & 6                       & 4317              & 6                   \\
Hgamm     & 4340         & 40                 & 4266          & 8                       & 4376              & 8                   \\
HeI3      & 4387         & 14                 & 4367          & 10                      & 4407              & 20                  \\
MnI2      & 4458         & 12                 & 4442          & 6                       & 4478              & 8                   \\
HeI4      & 4471         & 14                 & 4460          & 8                       & 4491              & 10                  \\
MnII      & 4481         & 8                  & 4473          & 8                       & 4501              & 20                  \\
FeI3      & 4490         & 20                 & 4470          & 20                      & 4510              & 16                  \\
FeI4      & 4532         & 12                 & 4520          & 6                       & 4552              & 12                  \\
FeI5      & 4592         & 24                 & 4572          & 14                      & 4610              & 8                   \\
FeI6      & 4669         & 10                 & 4649          & 10                      & 4679              & 10                  \\
FeI7      & 4787         & 8                  & 4777          & 12                      & 4797              & 6                   \\
Hbet      & 4861         & 40                 & 4815          & 8                       & 4905              & 8                   \\
HeI5      & 4922         & 14                 & 4902          & 14                      & 5042              & 20                  \\
HeI6      & 5016         & 20                 & 4996          & 12                      & 5032              & 6                   \\
FeI8      & 5079         & 8                  & 5059          & 6                       & 5091              & 8                   \\
FeII      & 5170         & 6                  & 5159          & 6                       & 5179              & 6                   \\
FeI9      & 5270         & 14                 & 5258          & 6                       & 5288              & 6                   \\
FeI10     & 5329         & 8                  & 5313          & 6                       & 5336              & 6                   \\
FeI11     & 5404         & 20                 & 5384          & 8                       & 5422              & 16                  \\
CaI2      & 5589         & 10                 & 5577          & 14                      & 5609              & 6                   \\
MgI       & 5711         & 10                 & 5691          & 20                      & 5725              & 10                  \\
HeI7      & 5876         & 8                  & 5860          & 6                       & 5884              & 6                   \\
NaI       & 5890         & 20                 & 5876          & 6                       & 5910              & 20                  \\
MnI3      & 6015         & 20                 & 5995          & 6                       & 6035              & 14                  \\
CaI3      & 6162         & 20                 & 6146          & 6                       & 6182              & 10                  \\
TiO1      & 6185         & 50                 & 6125          & 50                      & 6350              & 50                  \\
CaH2      & 6385         & 10                 & 6350          & 10                      & 6410              & 10                  \\

\end{tabular}
\end{table}

\begin{table}
\ContinuedFloat
\centering
\caption[]{(continued)}
\footnotesize
\setlength{\tabcolsep}{4pt}
\renewcommand{\arraystretch}{1.1}
\begin{tabular}{@{}lcccccc@{}}

Name & $\lambda_{FB}$ & $\Delta\lambda_{FB}$ & $\lambda_{BCB}$ & $\Delta\lambda_{BCB}$ & $\lambda_{RCB}$ & $\Delta\lambda_{RCB}$ \\
\hline
H$\alpha^a$    & 6563         & 50                 & 6505          & 10                      & 6610              & 10                  \\
HeI8      & 6678         & 10                 & 6658          & 18                      & 6698              & 8                   \\
Lith$^a$      & 6708         & 6                  & 6701          & 6                       & 6715              & 6                   \\
TiO3      & 6720         & 10                 & 6705          & 10                      & 6775              & 10                  \\
CaH3      & 6830         & 30                 & 6775          & 30                      & 7035              & 30                  \\
CaH1      & 6975         & 30                 & 6775          & 30                      & 7035              & 30                  \\
HeI9      & 7066         & 8                  & 7045          & 14                      & 7083              & 14                  \\
NaIT      & 8189         & 26                 & 8173          & 6                       & 8209              & 10                  \\
CaII1     & 8498         & 20                 & 8452          & 8                       & 8630              & 14                  \\
CaII2     & 8542         & 20                 & 8452          & 8                       & 8630              & 14                  \\
CaII3     & 8662         & 20                 & 8630          & 14                      & 8682              & 8                   \\
L\_TiO1b$^a$  & 4775         & 30                 & 4745          & 15                      & 4925              & 15                  \\
L\_TiO2b$^a$  & 4975         & 50                 & 4940          & 15                      & 5385              & 15                  \\
L\_TiO3b$^a$  & 5225         & 75                 & 4940          & 15                      & 5415              & 20                  \\
L\_TiO4b$^a$  & 5475         & 75                 & 5385          & 15                      & 5810              & 15                  \\
L\_TiO5b$^a$  & 5600         & 50                 & 5415          & 10                      & 5810              & 15                  \\
L\_TiO6b$^a$  & 5950         & 75                 & 5810          & 15                      & 6080              & 10                  \\
L\_TiO7b$^a$  & 6255         & 50                 & 6080          & 10                      & 6530              & 10                  \\
L\_TiO8b$^a$  & 6800         & 50                 & 6530          & 10                      & 7030              & 10                  \\
L\_TiO9b$^a$  & 7100         & 50                 & 7045          & 10                      & 7385              & 15                  \\
L\_TiO10b$^a$ & 7150         & 50                 & 7045          & 10                      & 7385              & 15                  \\
L\_VO1b$^a$   & 7460         & 100                & 7385          & 20                      & 7562              & 30                  \\
L\_VO2b$^a$   & 7940         & 100                & 7562          & 30                      & 8120              & 30                  \\
L\_VO3b$^a$   & 7840         & 100                & 7562          & 30                      & 8120              & 30                  \\
L\_VO4b$^a$   & 8500         & 120                & 8408          & 12                      & 8840              & 10                  \\
L\_VO5b$^a$   & 8675         & 150                & 8408          & 12                      & 8840              & 10                  \\
L\_VO6b$^a$   & 8880         & 40                 & 8840          & 10                      & 9045              & 10                  \\
C\_F1$^b$     & 4160         & 30                 & 4030          & 10                      & 4500              & 10                  \\
C\_F2$^b$     & 4380         & 10                 & 4030          & 10                      & 4500              & 10                  \\
C\_F3$^b$     & 4680         & 20                 & 4500          & 10                      & 4800              & 10                  \\
C\_F4$^b$     & 5100         & 30                 & 4800          & 10                      & 5250              & 10                  \\
C\_F5$^b$     & 5575         & 30                 & 5250          & 10                      & 5700              & 10                  \\
C\_F6$^b$     & 5980         & 30                 & 5700          & 10                      & 6250              & 10                  \\
C\_F7$^b$     & 6960         & 30                 & 6884          & 10                      & 7820              & 10                  \\
C\_F8$^b$     & 7116         & 30                 & 6884          & 10                      & 7820              & 10                  \\
C\_F9$^b$     & 7930         & 20                 & 7820          & 10                      & 8720              & 10                  \\
C\_F10$^b$    & 8100         & 20                 & 7820          & 10                      & 8720              & 10                  \\ 
\hline
\multicolumn{7}{l}{$^a$ Used in this work, }\\
\multicolumn{7}{l}{$^b$ used in \citet{Jaime2026}}\\

\end{tabular}
\end{table}


\begin{table*}
\centering
\caption{Summary of the datasets used throughout this work.}
\label{tab:datasets}

\begin{tabularx}{\textwidth}{
c
>{\raggedright\arraybackslash}p{2.0cm}
c
>{\raggedright\arraybackslash}X
>{\raggedright\arraybackslash}X
}
\toprule

Dataset &
Source &
N spectra &
Selection criteria &
Role \\

\midrule

Adapted synthetic library &
PHOENIX-based Göttingen library &
11,400 &
Synthetic spectra with $2300 \leq T_{eff} \leq 6000$ K (Sec.\ref{sec:phoenix}). &
GBM training, validation, and testing. \\

LAMOST DR10 &
\: -- -- &
11,441,011 &
All DR10 spectra. &
Initial spectroscopic sample. \\

$S_{0}$ sample &
LAMOST DR10 &
4,286,001 &
Quality cuts based on spectroscopic and Gaia criteria (Sec.\ref{sec:samples}). &
GBM temperature estimation. \\

LMS sample &
$S_{0}$ sample &
1,733,852 &
Objects with $2500 \leq T_{eff}^{\rm GBM} \leq 5100$ K (Sec. \ref{sec:regressor}). &
Low-mass star sample used for TTS identification. \\

Labeled sample &
LMS sample &
1,388 &
559 TTS and 829 non-TTS spectra (Sec.\ref{sec:samples}). &
Construction of the classification dataset. \\

Effective labeled sample &
Labeled sample &
1,374 &
552 TTS and 822 non-TTS spectra meeting classifier requirements (Sec.\ref{s:classifier}). &
Robust LR training, validation, and testing. \\

Production sample &
LMS sample &
1,732,478 &
LMS sample excluding the effective labeled sample (Sec.\ref{s:classifier}). &
Application of the robust LR classifier. \\

Classified sample &
Production sample &
1,719,380 &
Spectra with complete feature values for kernel PCA and robust LR. (Sec.\ref{s:classifier}). &
Probability-based TTS identification. \\

TTS candidate spectra &
Classified sample &
3,121 &
Spectra with $p\geq p_{thr}$ and gLiI EW recovered in $\geq90\%$ of iterations (Sec.\ref{s:classifier}). &
Final catalog of TTS candidate spectra.\\

\bottomrule
\end{tabularx}

\end{table*}



\bsp	
\label{lastpage}
\end{document}